\documentclass[showpacs,superscriptaddress,preprintnumbers,amsmath,amssymb,nofootinbib]{revtex4}

\usepackage{amsmath,amssymb,latexsym}
\usepackage{graphicx,bm}
\usepackage{dcolumn}    

\begin{document}
%

\newcommand{\sinc}{ {\mathrm{sinc}} }
\newcommand{\hf}{ {\hat{f}} }
\newcommand{\sqrtpi}{\sqrt{\pi}}

\def\H0{H_0}
\def\hatH{H}

\newcommand{\cA}{ a }

\title{ 
Probing anisotropies of gravitational-wave backgrounds  
\\
 with a space-based interferometer:  
\\
\it{ Geometric properties of   
  antenna patterns and their angular power
}
}

\author{Hideaki Kudoh} 
\email{kudoh_at_utap.phys.s.u-tokyo.ac.jp}
\affiliation{   
Department of Physics, The University of Tokyo, Tokyo 113-0033, Japan   
}

\author{Atsushi Taruya}
\email{ataruya_at_utap.phys.s.u-tokyo.ac.jp}
\affiliation{
Research Center for the Early Universe~(RESCEU), 
School of Science, The University of Tokyo, Tokyo 113-0033, Japan
}

\preprint{gr-qc/0411017, UTAP-504, RESCEU-35/04}
\pacs{04.30.-w, 04.80.Nn, 95.55.Ym, 95.30.Sf}

\begin{abstract}
We discuss the sensitivity to anisotropies of stochastic gravitational-wave backgrounds (GWBs) observed via space-based interferometer. In addition to the unresolved galactic binaries as the most promising GWB source of the planned Laser Interferometer Space Antenna (LISA), the extragalactic sources for GWBs might be detected in the future space missions. The anisotropies of the GWBs thus play a crucial role to discriminate various components of the GWBs. We study general features of antenna pattern sensitivity to the anisotropies of GWBs beyond the low-frequency approximation. We show that the sensitivity of space-based interferometer to GWBs is severely restricted by the data combinations and the symmetries of the detector configuration. The spherical harmonic analysis of the antenna pattern functions reveals that the angular power of the detector response increases with frequency and the detectable multipole moments with effective sensitivity
$h_{\rm eff}\sim 10^{-20}$Hz$^{-1/2}$ may reach 
$\ell\sim 8$--$10$ at $f\sim f_*=10$ mHz in the case of the single LISA detector. However, the cross correlation of optimal interferometric variables is blind to the monopole ($\ell=0$) intensity anisotropy, and 
also to the dipole ($\ell=1$) in some case, irrespective of the frequency band.
Besides, all the self-correlated signals are shown
to be blind to the odd multipole moments ($\ell=\mathrm{odd}$), independently of the 
frequency band.
\end{abstract}

\maketitle

\section{Introduction}
\label{sec:intro}

Space-based gravitational wave detectors retain many possibilities of 
providing access to new gravitational-wave sources that are not covered
by ground-based gravitational-wave detectors. The Laser Interferometer 
Space Antenna (LISA) is such a planned gravitational-wave observatory 
aimed at detecting and studying low-frequency gravitational waves in the 
band $0.1$ mHz $\sim$ $0.1$Hz. The constellation of the LISA and the
next generation detectors, e.g. DECIGO/BBO 
\cite{Seto:2001qf,Takahashi:2003wm,BBO:2003}, will consist of three 
spacecrafts keeping a triangle configuration.

Compared to ground-based detectors, a space-based gravitational-wave 
detector is characterized by many different features. For instance, the 
LISA introduces complications unknown to ground-based detectors, such as 
the complex signal and noise transfer functions. The complications block 
the analytical characterization of the detector~\cite{Vallisneri:2004,
Cornish:simulator,Cornish:2002rt,Merkowitz:2003uq,Merkowitz:2004st}, and 
in particular, the response of an interferometer becomes complicated for 
gravitational waves shorter than the arm length of the detector. Only in 
the low-frequency limit, the response of the LISA detector is very 
simplified. The three arms of the LISA function like a pair of two-arm 
detectors, and it is well known that the pair is equivalent in the low 
frequency limit to two $90^{\circ }$ interferometers which are rotated 
by $\pi/4$ with respect to each other (e.g. \cite{Cutler:1997ta}).

An important ingredient of a space-based detector is analysis of 
time-delayed combinations of data streams, which provide
laser-noise-free interferometric variable. The technique to synthesize
data streams is known as time delay interferometry (TDI) 
\cite{Tinto:1999yr,Armstrong:1999,Dhurandhar:2001kx,Prince:2002hp}.  
Several TDI signals, such as Michelson-like and Sagnac-like signals 
which are free from the laser frequency noise, will have different 
responses to 
secondary phase noise sources and to incoming gravitational waves. 
Starting with the original TDI observables for stationary-array 
combinations, the TDI observables have been developed until recently 
(see \cite{Krolak:2004xp} and references therein).

Space-based gravitational-wave detectors could be the most suitable 
devices to study and search for stochastic gravitational-wave signals. 
Examples of stochastic gravitational waves are those produced by large 
populations of Galactic \cite{Hils:1990hg,Bender:1997bc,Nelemans:2001hp}
and extra-galactic binaries \cite{Kosenko:1998,Schneider:2000sg,Farmer:2003pa} 
and 
a primordial gravitational-wave 
background produced by several cosmological mechanisms 
(see \cite{Maggiore:1999vm} for a review). 
Stochastic gravitational waves are expected to be anisotropic, and 
an important issue is to identify unambiguously the anisotropy to get 
insights into the origin and underlying physics of them.

A method to explore an anisotropy of gravitational-wave 
background has been recently proposed based on the time 
modulation of the single data stream 
\cite{Giampieri:1997,Giampieri:1997ie,Cornish:2001hg,Cornish:2002bh,
Ungarelli:2001xu} and/or the two data streams 
\cite{Cornish:2001hg,Seto:2004np,Allen:1997gp},  
which allow us to extract the individual coefficients of 
multipole moments related to a distribution of gravitational waves 
on the sky. Hence, provided all the coefficients of 
multipole moments observationally, one 
can, in principle, make the sky map of the gravitational-wave 
backgrounds \cite{Cornish:2001hg,Cornish:2002bh}. 
It was demonstrated in the low 
frequency limit that the LISA is blind to the whole set of odd multipole 
moments and sensitive only to monopole ($\ell=0$), quadrupole 
($\ell=2$) and octupole ($\ell=4$) anisotropy 
\cite{Ungarelli:2001xu,Cornish:2001hg}. 
Actually, the multipole moments $\ell=2, 4$ and their $m$ th harmonics 
of the galactic distribution of binaries would be observable with 
sufficiently high signal-to-noise ratios, except for some multipole 
harmonics \cite{Seto:2004np}. The restricted sensitivity to the
multipole moments is an immediate outcome of the low-frequency 
approximation, but what is the underlying physics that determines the 
limitation? As discussed in this paper, it is intimately associated 
with the geometric properties of the spacecraft configuration.

So far most of the works aimed at probing the anisotropy of
gravitational-wave background by means of space-based interferometers 
have been restricted to the low-frequency approximation. One 
reason is that a confusion gravitational-wave background formed by the 
superposition of many Galactic binaries comes in the low-frequency band 
of the LISA as the dominant source \cite{Hils:1990hg}. 
However there are many expected
sources of gravitational-wave backgrounds that spread outside the 
low-frequency region \cite{Kosenko:1998,Schneider:2000sg,
Farmer:2003pa,Schneider:2000sf,Enoki:2004ew}. Thus it is an interesting
problem to create maps of the gravitational-wave background in a very 
wide range of frequency. For that purpose we need to know the general 
response and properties of space-based interferometer over a wide range 
of frequency.

In this paper we are interesting in general features of response
function for space-based detectors.  The sensitivity of space-based 
detectors to multipole moments of a gravitational-wave distribution is 
in general restricted by symmetries of a response function independently 
of frequency. For example, symmetries of a response function tell us
that a self-correlated data is blind to the odd multipole moments of 
anisotropy irrespective of frequency band (Sec. \ref{sec:Parity of detector response}). Other interesting features 
independent of frequency can be also derived based on symmetries of a 
detector's response and geometric configuration of the spacecrafts.

The paper is organized as follows. 
After briefly reviewing the detection method of an anisotropy by the 
correlation analysis in the next section, 
detector response functions for space-based interferometers  
are given in Sec.~\ref{sec:Detector Response}. In 
Sec.~\ref{sec:Spherical harmonic analysis}, we develop spherical
harmonic analysis of antenna pattern functions and derive various 
fundamental properties of multipole moments. 
Based on those fundamental properties, in Sec.~\ref{sec:directional
sensitivity}, we examine the directional sensitivity of space
interferometer. Angular power and effective sensitivity curves are
discussed there, specifically focusing on the LISA detector. 
Section \ref{sec:Conclusion} concludes the paper with a brief summary. 
Below the speed of light is set equal to unity ($c=1$).

\section{detection of anisotropy through the correlation analysis}
\label{sec:detection of anisotropy}

We begin by discussing how one can prove the anisotropy of
gravitational-wave background based on the correlation analysis. 
A stochastic background of gravitational waves can be expressed 
as a random superposition of plane waves propagating along 
$\mathbf{\Omega}=-\textbf{n}$ direction 
with surfaces of constant phase $\xi(\bf{x})=t + {\bf n\cdot x}$. 
Then the metric perturbation in transverse-traceless gauge 
$\textbf{h}$ is expressed as: 
\begin{eqnarray}
 \textbf{h}(t,\textbf{x}) 
  = \sum_{A=+,\times} \int_{-\infty}^{\infty} df 
     \int d  \mathbf{\Omega} \,\, 
        e^{i\, 2\pi \, f \xi}
         \widetilde{h}_A(f,  \mathbf{\Omega} ) \,\,
        { \bf{e}}^A(  \mathbf{\Omega}),
\label{eq:h-stocastic gw}
\end{eqnarray}
where  $\int d \mathbf{\Omega}$ denotes an integral over the sphere and 
$\tilde{h}_A (-f)= \tilde{h}^*_A{(f)}$ are the Fourier amplitudes 
of the gravitational waves for each polarization mode.  
The Fourier amplitudes $h_A$ is assumed to be characterized by the 
Gaussian random process:  
\begin{eqnarray}
\left\langle  \widetilde{h}_A  (f,  {\bf{\Omega}}) \right\rangle
        &=& 0,
\cr 
\left\langle 
\widetilde{h}_A^*  (f,  {\bf{\Omega}})
{{\widetilde{h}}_{A'} }  (f',  {\bf{\Omega}}')
\right\rangle
&=&
 \frac{1}{2} \delta(f-f') 
\frac{ \delta^2 (  {\bf{\Omega}},  {\bf{\Omega}}')}{4\pi}
 \delta_{AA'} S_h(|f|, \,\mathbf{\Omega}), 
\label{eq:h*h gaussian process}
\end{eqnarray}
where $S_h(|f|,\, {\bf \Omega})$ is the power spectral density of 
gravitational waves.  The polarization tensors ${
\bf{e}}^A(\bf{\Omega})$ appearing in equation (\ref{eq:h-stocastic gw}) 
may be explicitly given as follows:   
\begin{eqnarray}
    {\bf e}^+(  {\bf{\Omega}}) 
    = {\bf u} \otimes {\bf u} -  {\bf v} \otimes {\bf v} , \quad 
    {\bf e}^\times(  {\bf{\Omega}})
    = {\bf u} \otimes {\bf v} +  {\bf v} \otimes {\bf u},\,\, 
\label{eq:polarization tensor}
\end{eqnarray}
where the unit vectors ${\bf u}$, ${\bf v}$ are expressed in 
an ecliptic coordinate as:  
\begin{eqnarray}
{\bf u} &=& \cos\theta_E \cos \phi_E {\bf x}
          + \cos\theta_E \sin \phi_E {\bf y}
          - \sin \theta_E {\bf z}, 
\cr
{\bf v} &=&   \sin \phi_E {\bf x} - \cos \phi_E {\bf y}, 
\cr
{\bf n} &=&   \sin\theta_E \cos \phi_E {\bf x} 
          + \sin\theta_E \sin \phi_E  {\bf y}
          +  \cos \theta_E {\bf z} = -  {\bf \Omega}.
\label{eq: u,v,n}
\end{eqnarray}

The detection of a gravitational-wave background is achieved through the 
correlation analysis of two data streams. 
The output signal for the detector $I$ denoted by $s_I(t)$ is described 
by a sum of the gravitational-wave signal $h_I(t)$ and the detector
noise $n_I(t)$: 
\begin{equation*}
s_I(t) = h_I(t)+ n_I(t). 
\end{equation*}
We assume that the noise $n_I(t)$ is treated as a Gaussian random process with zero mean and spectral density 
$S_{\rm n}(f)$:
\begin{eqnarray*}
\left\langle  \widetilde{n}_I (f) \right\rangle
        &=& 0,
\cr 
\left\langle 
\widetilde{n}_I^*  (f){{\widetilde{n}}_{J} }  (f')
\right\rangle
&=&
 \frac{1}{2} \delta(f-f')\,\delta_{IJ}\,\, S_{\rm n}(|f|).
\end{eqnarray*}
Here, we further assume that the noise correlation between the two independent detectors is neglected 
\footnote{In the case of the space interferometer, while the various data streams can be constructed combining the signals extracted from respective space crafts, most of them  are dominated by a correlated noise. 
Thus, the optimal data combinations which cancel the correlated noise are required to work with the correlation analysis. }.

On the other hand, in addition to the information of a gravitational-wave background, the output signal $h_I(t)$ contains the time variation of the detector response caused by the detector motion.  
For example, the rotation of the Earth sweeps the ground-based interferometer across the sky. As for the space interferometer, LISA, the antenna pattern sweeps over the sky as the LISA constellation orbits around the sun with period of one sidereal year. These effects induce the signal modulation, which can be used to extract the information of anisotropy of gravitational-wave backgrounds.

According to Ref.\cite{Allen:1997gp}, we introduce two time-scales, 
$\Delta T$ and $T_0$: the light travel time $\Delta T$ between the two detectors (space crafts) and the period of the detector motion $T_0$. 
Since $\Delta T\ll T_0$, it is possible to choose the averaging time scale $\tau$ as $\Delta T\ll \tau \ll T_0 $ appropriately. 
Then, one can safely employ the correlation analysis between two detectors as a function of time averaged over the period $\tau$. 
Keeping this situation in mind, the output signal $h_I(t)$ may be written as 
\begin{equation}
h_I(t) = \sum_{A=+,\times} \int_{-\infty}^{+\infty} df 
\int d \mathbf{\Omega} \,\,\textbf{D}_I( \mathbf{\Omega},f; t) \,\textbf{:}~
\textbf{e}^A( \mathbf{\Omega})\,\, \widetilde{h}_A(f, \mathbf{\Omega})\,\,
e^{i2\pi f\,\xi(x_I)},
\end{equation}
where the colon denotes the double contraction, i.e., 
$\textbf{D:\,e}=D_{ij}e^{ij}$ \cite{Cornish:2001qi}.   
The quantity $\textbf{D}$ is detector's response function, whose explicit expression will be presented in next section. 
Note that the response function depends on time due to the detector motion.

Provided the two output data sets, the correlation analysis is examined depending on the strategy of data analysis, i.e., self-correlation analysis only using the single data stream or cross-correlation analysis using the two {\it independent} data stream: 
\begin{eqnarray*}
  C(t)&\equiv&  \left\langle s_I(t)s_J(t) \right\rangle
    =     \left\langle h_I(t)h_J(t) 
        \right\rangle + \left\langle n_I(t)n_J(t) \right\rangle
\cr
    &=&  \int^{\infty}_{-\infty} \frac{df}{2}   \int \frac{d
    \mathbf{\Omega}}{4\pi}  S_h(|f|, \mathbf{\Omega})~ 
\mathcal{ F}_{IJ}^E(f,  \mathbf{\Omega};\,t)
         + \delta_{IJ} \int^{\infty}_{-\infty}  \frac{ df}{2} S_n(|f|), 
\end{eqnarray*}
where $\mathcal{F}^E_{IJ}$ is the antenna pattern function defined in an ecliptic coordinate, which is expressed in terms of detector's response function and an optimal filter: 
\begin{eqnarray}
&& \mathcal{F}^E_{IJ}(f, \mathbf{\Omega};\,t)=  
      e^{ i\, 2\pi f{\bf \Omega \cdot}(\textbf{x}_I -\textbf{x}_J)/L }
 \sum_{A=+,\times} 
F_I^{A*}(  \mathbf{\Omega},f;\,t) F_J^A( \mathbf{\Omega},f;\,t)
\, \widetilde{Q}(f)\,
\cr
&& F_I^A(  \mathbf{\Omega},f;\,t)  = 
{\bf D}_I( \mathbf{\Omega},f;\,t)\, \textbf{:} \,
\textbf{e} ^A (\mathbf{\Omega})
\label{eq:def_of_antenna}
\end{eqnarray}
with $\widetilde{Q}(f)$ being the Fourier transform of the optimal filter. 
Note that the phase factor 
$e^{i\,2\pi\,f\, \mathbf{\Omega}{\cdot}(\textbf{x}_I
-\textbf{x}_J)/L}$ arises due to the differences between the arrival time of the signal at each detector. The above expression implies that the time series data $C(t)$  as observable is given by the all-sky integral of the spectral density $S_h$, or luminosity distribution of gravitational waves convolving with the antenna pattern function. 
To see this more clearly, for the moment, we neglect the noise contribution and set the optimal filter as $\tilde{Q}(f)=1$. Keeping the assumption $\Delta T\ll t\ll T_0$, the detector output $C(t)$ is written as 
\begin{eqnarray}
   C(t) &=& \int_{-\infty}^{\infty} \frac{df}{2} \,\,
        \widetilde{C}(t,f) 
\cr
&=& \int_{-\infty}^{\infty} \frac{df}{2} \int 
\frac{d \mathbf{\Omega}}{4\pi} ~
         {S}_h(|f|, \mathbf{\Omega})\, \mathcal{F}^E_{IJ}(f, t,\mathbf{\Omega}).  
\label{eq:detector output C(t)}
\end{eqnarray}   
We then decompose the antenna pattern function 
$\mathcal{F}_{IJ}^E$ and the luminosity distribution into spherical harmonics in an ecliptic coordinate, i.e., sky-fixed frame. We have 
\begin{eqnarray}
    S_h(|f|,\, \mathbf{\Omega}) 
    = 
    \sum_{\ell m} [p^{E}_{\ell m} (f)]^* \,\,
        Y_{\ell m}^* ( \mathbf{\Omega}), 
\quad
    \mathcal{F}^E_{IJ}(f,\, \mathbf{\Omega};\,t)
    = 
    \sum_{\ell m} a_{\ell m}^E (f,t) \,\,
        Y_{\ell m} ( \mathbf{\Omega}). 
\label{eq:Y_lm expansion of S and F}
\end{eqnarray}
Substituting (\ref{eq:Y_lm expansion of S and F}) into 
(\ref{eq:detector output C(t)}) becomes 
\begin{eqnarray}
 \widetilde{C}(t,f) = \frac{1}{4\pi}\, \sum_{\ell m}\,\, 
\left[ p^E_{\ell m}(f) \right]^*  a^{E }_{\ell m}(f,\,t). 
\end{eqnarray}
Note that the time dependent multipole coefficient $a^{E}_{\ell m}$ appears due to the detector motion, which can be eliminated by further employing the harmonic expansion in detector's rest frame $(\theta,\phi)$. 
We denote the multipole coefficients of the antenna pattern in detector's rest frame by $a_{\ell m}$ (see 
Eq.[\ref{eq:a_lm_in_detector_frame}]). The transformation between the detector rest frame and the sky-fixed frame is described by a rotation matrix by the Euler angles $(\psi, \vartheta ,\varphi)$, whose explicit relation is expressed in terms of the Wigner $D$ matrices 
\cite{Allen:1997gp,Cornish:2002bh, Edmonds:1957}: 
\begin{eqnarray}
    a^E_{\ell m }(f,t)  = \sum_{n=-\ell}^\ell
    e^{-i \,n \,\psi}\, d^\ell_{nm}( \vartheta)\, e^{-i \,m \,\varphi}
\,\,a_{\ell n}(f). 
\label{eq:Wigner_formula}
\end{eqnarray}
Here the Euler rotation is defined to perform a sequence of rotation, starting with a rotation by $\psi$ about the original $z$ axis, followed by rotation by $\vartheta$ about the original $y$ axis, and ending with a rotation by $\varphi$ about the original $z$ axis. Note that the Euler rotation conserves the multipole moment $\ell$, but mixes $m$-th harmonics.

For illustration, let us envisage the orbital motion of the LISA constellation. The LISA orbital motion can be expressed by 
$\psi=-\omega t$, $\vartheta=-\pi/3$, $\varphi= \omega t$, where $\omega = 2\pi/T_0$ is LISA's orbital frequency ($T_0=1$ sidereal year). 
Since the antenna pattern function is periodic in time due to the orbital motion, one can naturally perform the Fourier transformation of the detector output by \cite{Giampieri:1997,Cornish:2001hg}: 
\begin{eqnarray*}
    \tilde{C}_k(f) =  \frac{1}{T_0} 
        \int^{T_0}_{0} dt~ e^{-ik \omega t}\,\, \tilde{C}(t,f). 
\end{eqnarray*}
Using the relation (\ref{eq:Wigner_formula}), we finally obtain:  
\begin{eqnarray}
    \tilde{C}_k(f) = 
        \frac{1}{4\pi}\,\,\sum_{\ell=0}^{\infty} \sum_{m=-\ell}^{\ell-k}
     [p_{\ell m}^E(f)]^*\,\, 
        d^{\ell}_{(m+k),m}\left(\vartheta \right) \,a_{\ell,(m+k)}(f) ,  
\label{eq:deconvolution}
\end{eqnarray}
for $k\ge 0$. The above equation shows how the detector output depends on the multipole coefficients $a_{\ell m}$ in detector's rest frame for a given luminosity distribution of a gravitational-wave background, 
$p_{\ell m}^E(f)$. Given the output data $\tilde{C}_k(f)$ experimentally, the task is to solve the linear system 
(\ref{eq:deconvolution}) with respect to $p_{\ell m}^E(f)$ if we know
the antenna pattern function.  As discussed by Cornish 
\cite{Cornish:2001hg}, this deconvolution problem is typically either over-constrained or under-constrained depending on the antenna pattern. 
In this sense, the understanding of the general properties of antenna pattern functions is primarily important and would shed light on the deconvolution problem. It might be further helpful to characterize the directional sensitivity to the sky map of the gravitational-wave background. 
The detailed investigation for the spherical harmonic analysis of antenna pattern will be presented in 
Sec.~\ref{sec:Spherical harmonic analysis}.  
Before developing the analysis, we briefly review the detector response functions for space interferometers.

\section{Detector response function for space interferometer}
\label{sec:Detector Response}

In this section, according to the treatment based on the coordinate-free 
approach in Ref. \cite{Cornish:2002rt,Cornish:2001bb},  we derive
various types of detector response functions for space interferometer, 
which will be used to analyze  the sensitivity to an anisotropy of 
gravitational-wave background.

\subsection{One-arm detector tensor}
\label{subsec:One-arm detector tensor}

Following the Doppler tracking calculations described in Ref. 
\cite{Cornish:2001qi}, the optical-path length between spacecraft $i$ 
and spacecraft $j$ is formally written as 
\begin{eqnarray}
\ell _{ij}(t_i) = \int^j_i \sqrt{ g_{\mu\nu} dx^\mu dx^\nu }, 
\end{eqnarray}
where $g_{\mu\nu}$ is spacetime metric.  
According to Ref. \cite{Cornish:2002rt}, the optical-path variation 
in presence of the gravitational waves is given by
\begin{eqnarray}
\delta \ell_{ij} (t_i)
= \ell_{ij}(t_i) 
\int^\infty_{-\infty} df\,\,\int\, d\mathbf{\Omega}\,\,\,
\textbf{D}(f,t_i, {\bf n})  \textbf{:} \tilde{\textbf{h}}
(f,\,\mathbf{\Omega})  
  e^{i\,2\pi\, f (t_i + {\bf n}  \cdot  \textbf{x}_i)}, 
\end{eqnarray}
where the one-arm detector tensor $\textbf{D}$ and the transfer function 
$\mathcal{T}$ are 
\begin{eqnarray}
    \textbf{D}(f,t_i, \textbf{n})
&=&
    \frac{1}{2} [ \textbf{r}_{ij} (t_i) \otimes  \textbf{r}_{ij} (t_i) ]
    \mathcal{T} (w,t_i, \textbf{n}),
    \label{eq:one-arm_response}
\\
    \mathcal{T} (f,t_i, \textbf{n})
&=& 
    \sinc \left[ \frac{ f }{2f_{ij}} [1 + \textbf{n} \cdot 
\textbf{r}_{ij} (t_i) ] \right]
    \exp\left[ i \frac{ f }{2f_{ij}} [1 + \textbf{n} \cdot 
\textbf{r}_{ij} (t_i) ] \right],
    \label{eq:transfer_one-arm}
\end{eqnarray}
where $\textbf{r}_{ij}(t_i)$ is the unit vector pointing from the space craft $i$ at the time of emission $t_i$ to the space craft $j$ at the time of reception $t_j$, i.e., 
$\textbf{r}_{ij}(t_i)=\{{\bf x}_j(t_j)-{\bf x}_i(t_i)\}/l_{ij}(t_i)$. 
The function  $\sinc(x)$ is defined by $\sinc(x)= \sin x/x $ and the variable $f_{ij}= [2\pi\ell_{ij}(t_j)]^{-1}$ means the characteristic transfer frequency.

\subsection{Detector response function}

The calculation of the one-arm detector tensor can be applied to derive 
the response function for a space interferometer via Doppler tracking
method. The constellation of the planned space interferometer, LISA and 
also the next generation detectors DECIGO/BBO constitutes three space 
crafts and each of them is separated in an equal arm length 
(Fig. \ref{fig:confgulation1}). Note cautiously that the detector arm 
length varies in time, mainly due to the intrinsic variation by the 
Keplerian motion of three spacecrafts and the tidal variation caused by 
the gravitational force of Solar System planets 
\cite{Bender:1998,Rubbo:2003ap}. 
The caveats concerning 
these effects have been already mentioned \cite{Cornish:2003tz} 
and their influences were 
recently investigated. 
As long as the low-frequency gravitational waves with
frequencies comparable or lower than the characteristic frequency 
$f_{ij}$ are concerned \cite{Cornish:2002rt,Rubbo:2003ap,Vecchio:2004ec}, 
the so-called {\it rigid adiabatic approximation}
\cite{Rubbo:2003ap}, in which the three space crafts rigidly orbit 
around the sun under keeping their configuration, really works.

Keeping these remarks in mind, we adopt the rigid adiabatic
approximation to give an analytic expression for response functions. 
For the sake of the brevity, we work with the static and the equal-arm 
limit of the detector response. 
In this case, one writes $L=l_{ij}$ and $f_*\equiv f_{ij}=1/(2\pi L)$. 
Thus, the interferometric signals combining with the six data streams 
can be generally expressed as function of 
\begin{equation}
\hat{f} \equiv \frac{f}{f_*}.
\label{eq:hat_f}
\end{equation}
Specifically, for LISA detector, the arm length is $L=5\times10^6$km, yielding 
$f_*\simeq 10$mHz.

Based on the configuration in Fig. \ref{fig:confgulation1}, 
a signal of Michelson interferometers extracted from the space craft $1$ is 
\cite{Cornish:2002rt,Cornish:2001bb}:
\begin{eqnarray}
h_{\scriptscriptstyle\rm M_1}(t) &=& \frac{1}{ 2L } 
   \Bigl[     \delta \ell_{12}(t-2L) 
             + \delta \ell_{21}(t-L) 
             - \delta \ell_{13}(t-2L) 
             - \delta \ell_{31}(t-L) 
    \Bigr]
\cr
  &=&  
  \int^\infty_{-\infty} df\,\,\int d\mathbf{\Omega}\,\,    
   \textbf{D}_{\scriptscriptstyle \rm M_1}
(\mathbf{\Omega},f) \textbf{:}  \widetilde{\textbf{h}}
        (f,\mathbf{\Omega})\,\,e^{i\,2\pi\, f \xi(x_1)  },
\label{eq:s^M-Michelson}
\end{eqnarray}
where $\textbf{D}_{\scriptscriptstyle \rm M_1}(\mathbf{\Omega},\,f)$ 
is the detector tensor. 
The explicit form of the detector tensor is given by 
\begin{eqnarray}
{\bf D}_{\scriptscriptstyle\rm M_1}(\mathbf{\Omega},f) 
&=& \frac{1}{2}\left\{ ({\bf a}\otimes{\bf a})\,
        {\cal T}_{\scriptscriptstyle\rm M}({\bf a}\cdot{\bf\Omega},f) 
 -    ({\bf c}\otimes{\bf c})\, {\cal T}_{\scriptscriptstyle\rm M}(-{\bf
     c}\cdot{\bf\Omega},f) \right\} ~~;
\cr
{\cal T}_{\scriptscriptstyle\rm M}({\bf u}\cdot{\bf\Omega}, f)
&=& 
 e^{-i \hf}
\left\{
   \sinc \left( \frac{ \hf(1-{\bf u} \cdot{\bf\Omega})}{2}  \right)
    e^{ - \frac{i}{2}\hf (1 +{\bf  u}\cdot{\bf\Omega}) } 
 + \sinc \left( \frac{ \hf(1+{\bf u}\cdot{\bf\Omega})}{2}  \right) 
    e^{ - \frac{i}{2}\hf( -1+{\bf u} \cdot {\bf\Omega})  }
\right\} .
\label{eq:D_{M}} 
\end{eqnarray}
The directional unit vectors for the three space crafts are denoted 
by $\mathbf{a}, \mathbf{b}, \mathbf{c} $ (Fig.~\ref{fig:confgulation1}).
Note that the above expression possesses the symmetry, i.e.,  
${\bf D}_{\scriptscriptstyle\rm M} (\mathbf{\Omega}, -f) = 
{\bf D}_{\scriptscriptstyle\rm M} (\mathbf{\Omega}, f)^*$.

Unfortunately, the simple Michelson interferometry with unequal armlengths does not cancel the laser frequency noise, which is thought to be one of the most dominant sources in the instrumental noises. 
Thus, the Michelson signal might not be a viable interferometric variable. Instead, a number of so-called TDI variables that cancel the laser frequency noise even when the armlengths are unequal have been proposed \citep[]{Armstrong:1999}. 
These signals are built by combining time-delayed Michelson signals so
as to reduce the overall laser frequency noise down to a level of other noises.
A particular example of a TDI variable is the X signal:  
\begin{eqnarray}
h_{\scriptscriptstyle\rm X_1} (t)
&=&  \frac{1}{4L} \Bigl[
        \delta\ell_{12}(t-2L) - \delta\ell_{12}(t-4L)
     +  \delta\ell_{21}(t- L) - \delta\ell_{21}(t-3L)
\nonumber
\\
&&     -  \delta\ell_{13}(t-2L) + \delta\ell_{13}(t-4L)
     -  \delta\ell_{31}(t- L) + \delta\ell_{31}(t-3L) 
     \Bigr].
\end{eqnarray}
This signal is expressed by a superposition of the Michelson signal, 
$s_{\scriptscriptstyle\rm  X_1} (t) =  
\frac{1}{2}[ s_1^{\scriptscriptstyle\rm M} (t) - 
s_1^{\scriptscriptstyle\rm M} (t-2L)]$. 
Thus, the detector tensor for the interferometer variable is 
$  {\bf D}_{\scriptscriptstyle\rm X}=  
\frac{1}{2} (1- e^{-2i \hat{f}}){\bf D}_{\scriptscriptstyle\rm M}$.

Other useful combination comes from comparing the phase of signals 
that are sent clockwise and counter-clockwise around the triangle. 
Such combination is named as Sagnac signal. 
The Sagnac signal extracted from the space craft $1$ is
\begin{eqnarray}
h_{\scriptscriptstyle\rm S_1}(t) &=&
        \frac{1}{3L} \Big[ \delta \ell_{13}(t-3L) 
                 +  \delta \ell_{32}(t-2L) +  \delta  \ell_{21}(t-L)
      -  \delta \ell_{12}(t-3L)-  \delta \ell_{23}(t-2L)
      -  \delta \ell_{31}(t-L) \Big] 
\cr
&=& 
  \int^\infty_{-\infty} df\,\,\int d\mathbf{\Omega}\,\,
{\bf D}_{\scriptscriptstyle\rm S_1} ({\bf\Omega},f) 
\,\textbf{:}\,
 \widetilde{\textbf{h}}
(f,\,{\bf\Omega})\,\,e^{2\pi i f \xi(x_1)},
\label{eq:s^S-Sagnac}
\end{eqnarray}
where the detector tensor ${\bf D}_{\scriptscriptstyle\rm S_1}$ 
is expressed as 
\begin{eqnarray}
{\bf D}_{\scriptscriptstyle\rm S_1}({\bf\Omega},f) 
&=&  \frac{1}{6}\left\{ ({\bf a}\otimes{\bf a})
   {\cal T}_{\rm a}(f) +
      ({\bf b}\otimes{\bf b})\, {\cal T}_{\rm b}(f)\right. 
 + \left.
    ({\bf c}\otimes{\bf c})\, {\cal T}_{\rm c}(f) \right\}~~;
\cr
{\cal T}_{\rm a}(f) &=&
e^{-3i \hf/2 } 
\left\{
e^{  - \frac{1}{2} i \hf  ( -2 + {\bf a}\cdot{\bf\Omega}  )  }
  \sinc \left[ \frac{ \hf}{2}\left(  1 + {\bf a}\cdot{\bf\Omega} \right) \right] 
- e^{- \frac{1}{2} i \hf( 2 + {\bf a}\cdot{\bf\Omega})  } 
  \sinc \left[ \frac{\hf}{2}  \left( 1-{\bf a}\cdot{\bf\Omega} \right) \right]
\right\},
\cr
{\cal T}_{\rm b}(f)
&=& 
 e^{- \frac{1}{2} i \hf (3+({\bf a}-{\bf c}) \cdot{\bf\Omega})}
\left\{ 
    \sinc
    \left[ \frac{ \hf}{2 }\left(1 + {\bf b}\cdot{\bf\Omega}\right) \right]  
  - \sinc \left[ \frac{\hf}{2 }\left(1-{\bf b}\cdot{\bf\Omega} \right)
               \right] 
\right\},
\cr
{\cal T}_{\rm c}(f) 
&=& 
e^{- \frac{3}{2}i\hf } 
\left\{
e^{- \frac{1}{2} i \hf (2 - {\bf c} \cdot {\bf\Omega})} 
  \sinc \left[ \frac{\hf}{2}  \left(1+{\bf c}\cdot{\bf\Omega}\right) \right] 
- 
e^{ \frac{1}{2} i\hf (2+{\bf c}\cdot{\bf\Omega})}
  \sinc \left[\frac{\hf}{2}\left( 1-{\bf c}\cdot{\bf\Omega}\right) \right]
\right\}.
\label{eq:D_{S}}
\end{eqnarray}
The three Sagnac signals extracted from the space crafts $1$, $2$ and $3$  
are often quoted as $\alpha, \beta$, and $\gamma$ in the literature 
(e.g., \cite{Armstrong:1999}). Combining these variables, 
a set of optimal data combinations free from the noise 
correlations is constructed\cite{Prince:2002hp} 
(see also \cite{Krolak:2004xp}): 
\begin{eqnarray}
h_{\rm A}&=& \frac{1}{\sqrt{2}}( h_{\scriptscriptstyle\rm S_3} 
- h_{\scriptscriptstyle\rm S_1} ),
\cr
h_{\rm E}&=& \frac{1}{\sqrt{6}}( h_{\scriptscriptstyle\rm S_1} 
- 2h_{\scriptscriptstyle\rm S_2} + h_{\scriptscriptstyle\rm S_3} ),
\cr
h_{\rm T}&=& \frac{1}{\sqrt{3}}( h_{\scriptscriptstyle\rm S_1} 
+  h_{\scriptscriptstyle\rm S_2} + h_{\scriptscriptstyle\rm S_3} ).  
\label{eq:def AET mode}
\end{eqnarray}

It is worthwhile to note that in the low-frequency limit $\hat{f}\ll1$, 
the detector tensor for the Michelson, the X and the Sagnac signal can 
be simply expressed as 
\begin{equation}
i\,\hat{f} \,{\bf D}_{\scriptscriptstyle\rm M_1} 
= {\bf D}_{\scriptscriptstyle\rm X_1} = \frac{3}{2}\,\,
{\bf D}_{\scriptscriptstyle\rm S_1}= \frac{i\hat{f}}{2}\,\,
( {\bf a}\otimes{\bf a}-{\bf c}\otimes{\bf c} )
+
\mathcal{O}(\hat{f}^2)
\label{eq:low_freq_response1}
\end{equation}
Using the above expression, the detector tensors for optimal combinations A, E and T respectively become
\begin{eqnarray}
{\bf D}_{\scriptscriptstyle\rm A}&=& \frac{i\,\hat{f}}{3\sqrt{2}}\,\,\,
\left( -{\bf a}\otimes{\bf a} -{\bf b}\otimes{\bf b} 
+2{\bf c}\otimes{\bf c}\right),
\nonumber\\
{\bf D}_{\scriptscriptstyle\rm E}&=& \frac{i\,\hat{f}}{\sqrt{6}}\,\,\,
\left( {\bf a}\otimes{\bf a} -{\bf b}\otimes{\bf b} \right),
\label{eq:low_freq_response2}\\
{\bf D}_{\scriptscriptstyle\rm T}&=& \frac{\,\hat{f}^2}{12\sqrt{3}}\,\,\,
\left\{ ({\bf a}\cdot{\bf \Omega})\,\,{\bf a}\otimes{\bf a} +
({\bf b}\cdot{\bf \Omega})\,\,{\bf b}\otimes{\bf b} + 
({\bf c}\cdot{\bf \Omega})\,\,{\bf c}\otimes{\bf c}  
\right\}.\nonumber
\end{eqnarray}
Note that the expression for detector tensor 
${\bf D}_{\scriptscriptstyle\rm T}$ is higher order in $\hat{f}$, compared to the other detector tensors. 
%
%
%
%
%
%
\section{Spherical harmonic analysis of antenna pattern function}
\label{sec:Spherical harmonic analysis}

The correlation analysis described 
in Sec.~\ref{sec:detection of anisotropy} reveals that the signal 
modulation induced by detector motion can be used to extract the 
information of the anisotropy of the gravitational-wave background. 
One important remark is that the map-making capability crucially 
depends on the antenna pattern and/or the detector response function 
in detector's rest frame. 
We then wish to clarify the relationship between the antenna pattern 
functions and the directional sensitivity to the gravitational-wave 
backgrounds. To investigate this issue, the spherical harmonic analysis 
of the antenna pattern function is employed and the general rules 
for multipole coefficients are derived based on the geometric 
properties of the antenna pattern.

\subsection{Angular power of antenna pattern function}

Similar to the expression (\ref{eq:def_of_antenna}), 
antenna pattern function $\mathcal{F}$ defined 
in detector's rest frame is written as 
\begin{eqnarray}
&& \mathcal{F}_{IJ}(f, \mathbf{\Omega})=  
      e^{ i \hat{f}  {\bf \Omega \cdot}(\textbf{x}_I -\textbf{x}_J) }
 \sum_{A=+,\times} 
F_I^{A*}(  \mathbf{\Omega},f) F_J^A(\mathbf{\Omega},f )
\, \widetilde{Q}(f)\,
\cr
&& F_I^A(\mathbf{\Omega},f )  = {\bf D}_I( \mathbf{\Omega},f) \textbf{:} 
\textbf{e} ^A ( \mathbf{\Omega}).
\label{eq:def_of_antenna 2}
\end{eqnarray}
The multipole coefficient $a_{\ell m}$ for (\ref{eq:def_of_antenna 2}) is  
\begin{equation}
a_{\ell m}(\hat{f}) = \int_0^{\pi}d\theta \int_0^{2\pi} d\phi 
\sin \theta 
\, Y_{\ell m}^* (\theta,\phi)
\, \mathcal{F} (\hat{f},\theta,\phi)
\label{eq:a_lm_in_detector_frame}
\end{equation}
with $\hat{f}$ being defined in (\ref{eq:hat_f}). We are primarily
concerned with how the directional sensitivity depends on the choice of 
the interferometric variables. For this purpose, the optimal filter 
$\widetilde{Q}(f)$ appearing in the antenna pattern function 
(\ref{eq:def_of_antenna}) is ignored hereafter. Using the fact that the 
relations $ {\mathcal F}^*(\hat{f},{\bf\Omega}) 
 = {\mathcal F} (-\hat{f},{\bf\Omega}) 
 = {\mathcal F} (\hat{f}, -{\bf\Omega})$ always hold\footnote{The last equality comes directly from the properties of the respective detector tensors, and thus it holds only among the same types of TDI variables, e.g., ${\mathcal F}_{S_I S_J}$ and ${\mathcal F}_{M_I M_J}$. },  
one obtains 
\begin{eqnarray}
a_{\ell, m} (\hat{f},t) 
= (-1)^m a_{\ell, -m}^* (-\hat{f},t)
= (-1)^{m+\ell} a_{\ell, -m}^* (\hat{f},t). 
\label{eq:parity relation of a_l-m}
\end{eqnarray}
where we used 
$Y_{\ell -m}(\theta, \phi) = (-1)^m Y_{\ell m}^*(\theta, \phi)$.

\begin{figure}
\begin{center}
\includegraphics[width=7cm,clip]{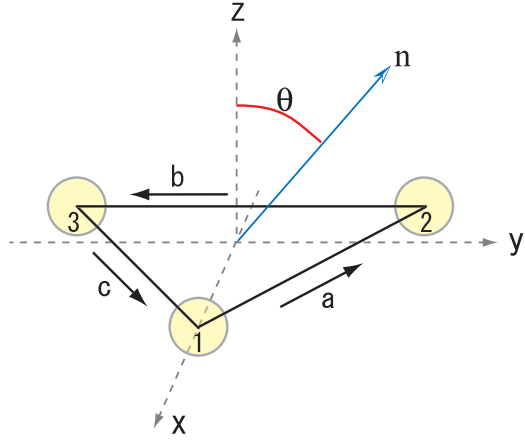}
\caption{
\label{fig:confgulation1}
Configuration of the spacecraft constellation in detector's rest frame.
}
\end{center}
\end{figure}

Here and in what follows, we consider the detector configuration in a specific coordinate system shown in Fig. \ref{fig:confgulation1} to calculate the multipole coefficients. Unless otherwise stated, the unit vectors ${\bf a}, {\bf b}$ and ${\bf c}$ 
are specifically chosen as:  
\begin{eqnarray}
{\bf{a}} = -\frac{\sqrt{3}}{2} {\bf{x}} + \frac{1}{2} {\bf{y}}, 
\quad
{\bf{b}} = - {\bf{y}}, 
\quad
{\bf{c}} =   \frac{\sqrt{3}}{2} {\bf{x}} + \frac{1}{2} {\bf{y}}. 
\label{eq: a,b,c}
\end{eqnarray}
While the explicit form of the multipole coefficients $a_{\ell m}$
depends on the coordinate system (\ref{eq: a,b,c}), a convenient
quantity invariant under a Euler rotation of the coordinate system can 
be exploited: 
\begin{eqnarray}
    \sigma_\ell^2 (\hat{f}) = \frac{1}{2\ell+1}\sum_{m=-\ell}^{\ell} 
 |a_{\ell m}(\hat{f})|^2, 
\label{eq: sigma_ell1}
\end{eqnarray}
which characterizes the contribution of $\ell$-th moment to the antenna 
pattern function. Thus, under the rigid adiabatic approximation, the 
angular power of the antenna pattern in the ecliptic frame is equivalent 
to that in detector's rest frame:  
\begin{eqnarray}
 \sum_{m } 
 |a_{\ell m}(\hat{f})|^2 = 
 \sum_{m } 
 |a_{\ell m}^{E}(\hat{f},t)|^2. 
\label{eq: sigma_ell2}
\end{eqnarray}
We use this coordinate invariant quantity to quantify the directional 
sensitivity of the antenna pattern.

\subsection{Low-frequency limit}
\label{sec:low_freq}

Consider first the simplest case, $\hat{f}=f/f_*\ll1$. 
In this case, only the $\ell=0$, $2$ and $4$ moments for antenna pattern 
function become non-vanishing. This is mostly general except for the
fully symmetrized signals such as $T$-variable.

In the low-frequency limit, the detector response functions derived in 
previous section generally becomes of the form 
(see Eqs.[\ref{eq:low_freq_response1}][\ref{eq:low_freq_response2}]): 
\begin{equation}
{\bf D}\longrightarrow  {\cal T}_{\rm a} ({\bf a}\otimes{\bf a}) 
        + {\cal T}_{\rm b} ({\bf b}\otimes{\bf b}) 
        + {\cal T}_{\rm c} ({\bf c}\otimes{\bf c}),  
\label{eq:general_form_low-freq}
\end{equation}
except for the $T$-variable. While the factors ${\cal T}_{\rm a,b,c}$ 
may be written as functions of 
frequency, they do not depend on the directional angle ${\bf\Omega}$.   
Thus, the detector tensor ${\bf D}$ loses 
the directional dependence. This means that 
the directional dependence of 
the antenna pattern function $\mathcal{F}$ arises 
only through the polarization tensor, ${e}^{\rm A}({\bf\Omega})$.  
Since the polarization tensor is described by the quadrature of 
direction vectors ${\bf u}$ and ${\bf v}$ (Eq.[\ref{eq: u,v,n}]), 
the antenna pattern can be generally written as the forth order 
polynomials of $(\cos\theta,\sin\theta)$ and $(\cos\phi,\sin\phi)$. 
For example, the low-frequency limit of the self-correlated signal 
for Michelson and Sagnac interferometries extracted from the spacecraft 
$1$ is 
\begin{equation}
\hat{f}^2\,\,{\mathcal F}_{\scriptscriptstyle\rm M_1M_1}
=\frac{9}{4} {\mathcal F}_{\scriptscriptstyle\rm S_1S_1}
~~\stackrel{\hat{f}\ll 1}{\longrightarrow}~~
\frac{3 }{4}
\left[
 \frac{1}{4}  (3 + \cos4\phi)\cos^2 \theta
 + (1 + \cos^4\theta) \cos^2\phi \sin^2\phi 
\right] \hat{f}^2
+{\mathcal O}(\hat{f}^4). 
\label{eq:antenna_self_correlation_low_freq}
\end{equation}
Applying the spherical harmonic expansion (\ref{eq:a_lm_in_detector_frame}), non-vanishing components of the multipole coefficients become 
\begin{eqnarray}
&&a_{00}^{\scriptscriptstyle\rm M_1M_1}=\frac{3\sqrt{\pi}}{5},
\quad
a_{20}^{\scriptscriptstyle\rm M_1M_1 }=\frac{6}{7}\sqrt{\frac{\pi}{5}},
\quad
a_{40}^{\scriptscriptstyle\rm M_1M_1 }=\frac{\sqrt{\pi}}{70},
\quad
a_{4,\pm4}^{\scriptscriptstyle\rm M_1M_1 } =-\frac{1}{2}\sqrt{\frac{\pi}{70}}.
\nonumber
\end{eqnarray}
The coordinate-free quantity $\sigma_l(\hat{f})$ is thus evaluated as 
\begin{eqnarray}
\sigma_0^{\scriptscriptstyle\rm M_1M_1 }=\frac{3\sqrt{\pi}}{5}, 
\quad
\sigma_2^{\scriptscriptstyle\rm M_1M_1 }=\frac{6\sqrt{\pi}}{35}, 
\quad
\sigma_4^{\scriptscriptstyle\rm M_1M_1 }=\frac{\sqrt{\pi}}{35}
\label{eq:sigma_self_michelson_low_freq}
\end{eqnarray}
for self-correlated Michelson signals. The angular power of 
self-correlated Sagnac signals are related to that of the Michelson 
signals by $\sigma_{\ell}^{\scriptscriptstyle S_1S_1}= 
(4/9)\hat{f}^2\,\,\sigma_{\ell}^{\scriptscriptstyle M_1M_1}$.  
As for the cross-correlated signal extracted from $1$ and $2$, the 
antenna pattern function is explicitly written as 
\begin{equation}
\hat{f}^2\,\,{\mathcal F}_{\scriptscriptstyle\rm M_1M_2}
=\frac{9}{4} {\mathcal F}_{\scriptscriptstyle\rm S_1S_2}
~~\stackrel{\hat{f}\ll 1}{\longrightarrow}~~
\frac{-3}{512}
\left[
 35 + 28 \cos 2\theta +\cos 4\theta 
 - 8 (\cos4\phi + \sqrt{3} \sin 4\phi)  \sin^4 \theta  
\right] \hat{f}^2
+{\mathcal O}(\hat{f}^3). 
\label{eq:antenna_cross_correlation_low_freq}
\end{equation}
We then obtain the non-vanishing multipole coefficients:  
\begin{eqnarray}
a_{00}^{\scriptscriptstyle\rm M_1M_2}=-\frac{3\sqrt{\pi}}{10}, 
\quad
a_{20}^{\scriptscriptstyle\rm M_1M_2}=-\frac{3}{7}\sqrt{\frac{\pi}{5}}, 
\quad
a_{40}^{\scriptscriptstyle\rm M_1M_2}=-\frac{\sqrt{\pi}}{140},
\quad
a_{4,\pm4}^{\scriptscriptstyle\rm M_1M_2}=
\frac{1\mp i\sqrt{3}}{4}\sqrt{\frac{\pi}{70}}, 
\nonumber
\end{eqnarray}
Correspondingly, the invariant quantity $\sigma_{\ell}$ becomes 
\begin{eqnarray}
\sigma_0^{\scriptscriptstyle\rm M_1M_2}=\frac{3\sqrt{\pi}}{10}, 
\quad
\sigma_2^{\scriptscriptstyle\rm M_1M_2}=\frac{3\sqrt{\pi}}{35}, 
\quad
\sigma_4^{\scriptscriptstyle\rm M_1M_2}=\frac{1}{140}\sqrt{\frac{47\pi}{3}}
\label{eq:sigma_cross_michelson_low_freq}
\end{eqnarray}
for Michelson signals and the same relation 
$\sigma_{\ell}^{\scriptscriptstyle\rm S_1S_2}= 
(4/9)\hat{f}^2\,\,\sigma_{\ell}^{\scriptscriptstyle\rm M_1M_2}$ holds for cross-correlated Sagnac signals.

The above examples show that the multipole coefficients 
$a_{\ell m}$ higher than $\ell=4$ vanish at the leading order in 
$\hat{f}$. Further, the lower multipole moments $\ell=1$ and $3$ also vanish because the antenna pattern ${\mathcal F}({\bf\Omega})$ is even function of $(\theta,\phi)$. 
This is irrespective of the choice of the coordinate system. 
Indeed, the same properties hold for the optimal combinations $A$, $E$ and $T$, since these are constructed from the linear combination of 
Sagnac variables.

In Appendix \ref{appendix:1}, the angular power of the optimal combinations are calculated analytically up to the forth order in $\hat{f}$. It is shown that the lowest order calculation for self-correlated signal $\sigma_{\ell}^{\scriptscriptstyle\rm AA}$ exactly coincides with that for $\sigma_{\ell}^{\scriptscriptstyle\rm EE}$, which is related to the self-correlated Michelson signal as $\sigma_{\ell}^{\scriptscriptstyle\rm AA}=\sigma_{\ell}^{\scriptscriptstyle\rm EE}=
(2/3) \hat{f}^2 \sigma_{\ell}^{\scriptscriptstyle\rm M_1M_1}$. 
On the other hand, the lowest order contribution to the self-correlated signal for $T$-variable becomes vanishing due to the symmetric combination of 
Sagnac variables. That is, the higher-order contribution of 
${\cal O}(\hat{f}^4)$ terms becomes dominant in the angular power of antenna pattern. 
While the resultant non-vanishing components for 
$\sigma_{\ell}^{\scriptscriptstyle\rm TT}$ are $\ell=0,~4$ and $6$, 
it turns out that the dominant noise contribution for 
$T$-variable appears as ${\cal O}(\hat{f}^2)$ 
(see Sec.\ref{subsec:sensitivity}). 
Therefore, the self-correlated signal for $T$-variable is dominated by 
the instrumental noise in the low-frequency limit and the 
gravitational-wave signal could not be resolved 
\cite{Armstrong:1999,Prince:2002hp,Cornish:2001bb}.

\subsection{Parity invariance in antenna pattern}
\label{sec:Parity of detector response}

Apart from the low-frequency limit as simple limiting approximation, 
the analytical calculation of $\sigma_{\ell}$ becomes intractable 
and the perturbative expansion for $\hat{f}$ generally breaks down. 
In contrast to the ground-based detectors, the difficulty in the space 
interferometers arises from the transfer function $\mathcal{T}$ that 
appears in equation (\ref{eq:one-arm_response}), which explicitly
exhibits both the frequency and the angular dependences. 
Thus, to evaluate the directional sensitivity, the numerical treatment 
is required for spherical harmonic analysis. 
Nevertheless, some important properties in the multipole coefficients 
of an antenna pattern can be still drawn analytically, 
from symmetric properties of an antenna pattern, which is closely-linked 
with the parity invariance of a detector response.

Let us introduce two operators, $\mathbb{Q}: \theta \to \pi - \theta$ and 
$\mathbb{P}: \phi \to \phi + \pi$. 
A composite operator $\mathbb{P\,Q}$ represents the parity transformation. 
The transformation properties of the spherical harmonics 
$Y_{\ell m}(\theta,\phi)$ for the operators are 
$\mathbb{P}\,Y_{\ell m} =  (-1)^m  Y_{\ell m}$ and 
$\mathbb{Q}\,Y_{\ell m} =  (-1)^{\ell+m} Y_{\ell m}$, 
and so the parity transformation property is 
$\mathbb{P\,Q}\,Y_{\ell m} = (-1)^{\ell} Y_{\ell m} $. 
If the response function is parity invariant, 
Eq. (\ref{eq:a_lm_in_detector_frame}) can be written as 
\begin{eqnarray}
  a_{\ell m} 
  &=&
    \int^{\pi}_{0} d\theta \sin \theta 
 \int^\pi_0 d\phi~Y^*_{\ell m} [ \mathcal{F}   +(-1)^{\ell} \,
\mathbb{P\,Q}\, \mathcal{F} ].  
\end{eqnarray} 
Then we see that $a_{\ell m}$ becomes vanishing for all odd multipoles
($\ell=$ odd) if a response function is parity invariant. 
 A similar argument holds for the respective operators 
$\mathbb{P}$ and $\mathbb{Q}$. If a detector response function is invariant 
under the operation $\mathbb{P}$, then $a_{\ell m}$ vanishes for $m=$odd 
moment.  For response function invariant under the operation 
$\mathbb{Q}$, the multipole coefficients $a_{\ell m}$ vanish when 
$\ell+m=\mathrm{odd}$. 
Here, we summarize these results\footnote{
There is another interesting property of the multipole coefficient.
Eq. (\ref{eq:parity relation of a_l-m}) tells us that 
the multipole coefficients of $\ell=\mathrm{even} ~(\mathrm{odd})$ 
modes are even (odd) functions of $\hat{f}$:  
$$
a_{\ell m} (-\hat{f}) 
= (-1)^{\ell} a_{\ell  m} (\hat{f}). 
$$
We will see this property explicitly through the low-frequency 
approximation in the following sections.
}: 
\begin{subequations}
\label{eq:a_lm rule}
\begin{alignat}{4}
 a_{\ell m} &=0 ~~ \mathrm{~for~}   
~&\mathbb{P}\,\mathcal{F} = \mathcal{F},\quad  &
~&m =\mathrm{odd},
\label{eq:a_lm_rule1}
\\  
 a_{\ell m} &=0 ~~ \mathrm{~for~}  
~&\mathbb{Q}\,\mathcal{F} = \mathcal{F},\quad  &
~& \ell+ m =\mathrm{odd}, 
\label{eq:a_lm_rule2}
\\  
 a_{\ell m} &=0 ~~ \mathrm{~for~} 
~&\mathbb{Q\,P}\,\mathcal{F} =\mathcal{F},\quad  &
~&\ell =\mathrm{odd}.
\label{eq:a_lm_rule3} 
\end{alignat}
\end{subequations}

Note that while the symmetric property of the antenna pattern itself is
a coordinate-free notion, the results presented in equations 
(\ref{eq:a_lm_rule1}) and (\ref{eq:a_lm_rule2}) depend on a choice of 
the coordinate system, since the mode $m$ can be mixed by the Euler 
rotation. On the other hand, the property (\ref{eq:a_lm_rule3}) that 
only depends on $\ell$ preserves under the Euler rotation.

Keeping this remark in mind, based on the specific configuration 
and the coordinate system shown in Fig.\ref{fig:confgulation1} and 
Eq. (\ref{eq: a,b,c}), several useful formulae 
related to the parity transformation are derived 
in Appendix \ref{appendix:parity trans}. 
Using these results, 
one finds that the antenna pattern 
functions for the self-correlated signals of 
Michelson, Sagnac and the optimal TDI variables are 
invariant under the following three transformations:  
\begin{alignat}{4}
\mathbb{Q\,P}~\mathcal{F}_{II} &= \mathcal{F}_{II},
\quad &
\mathbb{P}~\mathcal{F}_{II} &= \mathcal{F}_{II} , 
\quad  &
\mathbb{Q}~\mathcal{F}_{II} &= \mathcal{F}_{II}.  
\quad 
(I={\rm M}_i,\,{\rm S}_i,\,{\rm A,\,E,\,T}) &
\label{eq:QP,P,Q for self-correlation}
\end{alignat}
Thus, the multipole moments of antenna pattern functions for 
self-correlated signals follow the rule (\ref{eq:a_lm rule}).
 This is generally true in detector's rest frame under both the 
static and the equal-arm length limit. 
The antenna pattern function for the cross-correlated Sagnac signals obeys 
\begin{alignat}{4}
         \mathbb{Q\,P}~\mathcal{F}_{IJ}
         & =  \left[ \mathcal{F}_{IJ}\right]^*, \quad &
         \mathbb{P}~\mathcal{F}_{IJ} 
         & =  \left[ \mathcal{F}_{IJ} \right]^*, \quad &
         \mathbb{Q}~\mathcal{F}_{IJ} 
         & =  \mathcal{F}_{IJ}   \quad &
(I,J={\rm S}_i).
\label{eq:QP,P,Q for cross-correlation}
\end{alignat}

Several remarks concerning the optimal combinations 
are in order at this point. First recall that the 
antenna pattern functions constructed
from the signals A,\,E and T can be represented by a sum of the 
self-correlated and the cross-correlated Sagnac signals 
[see Eqs.(\ref{eq:Antenna, AA,EE,TT})(\ref{eq:Antenna, AE,AT,ET})].  
Thus, owing to the fact (\ref{eq:QP,P,Q for cross-correlation}) and the 
property  $[\mathcal{F}_{\scriptscriptstyle {\rm S}_i{\rm S}_j}]^* 
= \mathcal{F}_{\scriptscriptstyle{\rm S}_j {\rm S}_i}$, 
it can be shown that the 
property (\ref{eq:QP,P,Q for self-correlation}) holds for the 
self-correlated signals $\mathcal{F}_{\scriptscriptstyle\rm AA} $, 
$\mathcal{F}_{\scriptscriptstyle\rm EE}$ 
and $\mathcal{F}_{\scriptscriptstyle\rm TT}$, while 
the cross-correlated signals $\mathcal{F}_{\scriptscriptstyle\rm AE}$, 
$\mathcal{F}_{\scriptscriptstyle\rm AT}$ 
and $\mathcal{F}_{\scriptscriptstyle\rm ET}$ 
only possess the symmetry,  
$\mathbb{Q}\,\mathcal{F}=\mathcal{F}$. Hence, the 
cross-correlated data may generally contain the $\ell=\mathrm{odd}$ moments. 
Note, however, that in the low-frequency limit, 
the appreciable multipoles are the $l=0$, $2$ and $4$ modes. 
This readily implies that the contribution of the $\ell=\mathrm{odd}$ 
modes becomes significant at $\hat{f}=f/f_*\sim1$, which will be 
explicitly shown in next Section 
\ref{sec:directional sensitivity} [see angular power in 
Fig.~\ref{fig:sigma_optimal_TDIs} and effective sensitivity curves in 
Fig.~\ref{fig:h_eff_optimal_TDIs} ].

\subsection{Geometric relation between optimal combinations of TDIs}
\label{subsec:geometric_relation}

In a specific case with signals constructed from the optimal combinations (A,\,E,\,T),  a further important property is obtained combining with the geometric relationship among the three spacecrafts.

Let us start with the property of the Wigner $D$ matrices 
(\ref{eq:Wigner_formula}). For specific choice of the angles  
$\vartheta=0$ and $\vartheta=\pi$, the Wigner $D$ matrix becomes $d^\ell_{nm} (0)=\delta_{nm}$ and
$d^\ell_{nm} (\pi) = (-1)^{\ell+m}\delta_{n,-m}$, respectively \cite{Edmonds:1957}. 
Thus a rotation $\psi$ with $\vartheta=\varphi=0$ transforms the coefficients $a_{\ell m}$ to $a_{\ell m}'$ as 
\begin{eqnarray}
   a_{\ell m}' = e^{-i m \psi} a_{\ell m}. \quad  (\vartheta=\varphi=0)
\label{eq:a_{lm}-theta=0}
\end{eqnarray}
On the other hand, a rotation $\psi$ followed by the rotation 
$\vartheta=\varphi=\pi$ transforms the coefficients as 
\begin{eqnarray}
   a_{\ell m}' = (-1)^m e^{i m \psi} [ a_{\ell m} ]^* . 
\quad (\vartheta=\varphi=\pi)
\label{eq:a_{lm}-theta=pi}
\end{eqnarray}

From the spacecraft constellation shown in Fig. \ref{fig:confgulation1}, the antenna pattern functions of the self-correlated Sagnac signals 
$\mathcal{F}_{\scriptscriptstyle\rm S_{2}S_2}$ and 
$\mathcal{F}_{\scriptscriptstyle\rm S_{3}S_3}$ are 
related to $\mathcal{F}_{\scriptscriptstyle\rm S_1S_1}$ 
by the Euler rotation angles 
$\psi=2\pi/3$ and $\psi=4\pi/3$ ($\vartheta =\varphi=0$), respectively. 
Similarly, all multipole coefficients of cross-correlated Sagnac signals are related to $a_{\ell m}^{\scriptscriptstyle\rm S_1S_2}$ as indicated by 
(\ref{eq:a_{lm}-theta=0}) and (\ref{eq:a_{lm}-theta=pi}). 
These relations are summarized as follows:
\begin{eqnarray}
\begin{array}{llll}
a_{\ell m}^{\scriptscriptstyle\rm S_2S_2} = 
e^{- i m \delta} a_{\ell m}^{\scriptscriptstyle\rm S_1S_1},&
a_{\ell m}^{\scriptscriptstyle\rm S_3S_3} = 
e^{-2i m \delta} a_{\ell m}^{\scriptscriptstyle\rm S_1S_1},&
\\
\\
a_{\ell m}^{\scriptscriptstyle\rm S_2S_3} = 
e^{- i m \delta} a_{\ell m}^{\scriptscriptstyle\rm S_1S_2},& 
a_{\ell m}^{\scriptscriptstyle\rm S_3S_1} = 
e^{-2i m \delta} a_{\ell m}^{\scriptscriptstyle\rm S_1S_2},&
\\
\\
a_{\ell m}^{\scriptscriptstyle\rm S_1S_3} = 
(-1)^m [a_{\ell m}^{\scriptscriptstyle\rm S_1S_2}]^*,  & 
a_{\ell m}^{\scriptscriptstyle\rm S_2S_1} = 
(-1)^m e^{-4i m \delta} [a_{\ell m}^{\scriptscriptstyle\rm S_1S_2}]^*, &
a_{\ell m}^{\scriptscriptstyle\rm S_3S_2} = 
(-1)^m e^{-2i m \delta} [a_{\ell m}^{\scriptscriptstyle\rm S_1S_2}]^*,   &
\end{array}
\label{eq:Rotation a_{em}-self-cross}
\end{eqnarray}
where $\delta \equiv 2\pi/3$, and we have used (\ref{eq:parity relation of a_l-m}). 
This means that the antenna patterns for all the possible combinations of 
$(A,\,E,\,T)$ can be represented by a sum of the primary multipole moments of the Sagnac signals,  
$a_{\ell m}^{\scriptscriptstyle\rm S_1S_1}$ and 
$a_{\ell m}^{\scriptscriptstyle\rm S_1S_2}$. In this sense, the optimal combinations of TDIs are not strictly independent.

Based on the relations (\ref{eq:Rotation a_{em}-self-cross}), with a help of the expressions (\ref{eq:Antenna, AA,EE,TT}), the multipole moments for self-correlated signals 
$\mathcal{F}_{\scriptscriptstyle\rm AA}$ and $\mathcal{F}_{\scriptscriptstyle\rm EE}$ can be rewritten with 
\begin{eqnarray}
a_{\ell m}^{\scriptscriptstyle\rm AA}&=& C_{m}^{(1)}\,e^{-i\,m\delta}\,
a_{\ell m}^{\scriptscriptstyle\rm S_1S_1}- C_{m}^{(2)}
\left\{e^{-i\, 2m\delta}\,a_{\ell m}^{\scriptscriptstyle\rm S_1S_2}
+(-1)^m\, [a_{\ell m}^{\scriptscriptstyle\rm S_1S_2}]^*\right\},
\nonumber 
\\
a_{\ell m}^{\scriptscriptstyle\rm EE}&=& 
D_m^{(1)}\,e^{-i\,m\delta}\,
a_{\ell m}^{\scriptscriptstyle\rm S_1S_1}- D_m^{(2)}\,
\left\{e^{-i\, 2m\delta}\,a_{\ell m}^{\scriptscriptstyle\rm S_1S_2}
+(-1)^m\, [a_{\ell m}^{\scriptscriptstyle\rm S_1S_2}]^*\right\},
\label{eq:a^AA-a^EE}
\end{eqnarray}
where the coefficients $C_m^{(i)}$ and $D_m^{(i)}$ become
\begin{eqnarray}
    C_m^{(1)}=\cos({m\delta}),\quad
    C_m^{(2)} = \frac{1}{2},\quad
    D_m^{(1)}=\frac{1}{3}\left\{2+\cos({m\delta})\right\},\quad
    D_m^{(2)}=\frac{1}{6}\left\{4\cos({m\delta})-1\right\}.
    \nonumber
\end{eqnarray}
Now recall from the properties (\ref{eq:a_lm rule}) and (\ref{eq:QP,P,Q for self-correlation}) that the non-vanishing components of the multipole moments (\ref{eq:a^AA-a^EE}) are the $\ell=\mbox{even}$ and 
$m=\mbox{even}$ modes. Then the comparison between the coefficients $C_m^{(i)}$ and 
$D_m^{(i)}$ leads to the relation 
$a_{\ell m}^{\scriptscriptstyle\rm AA}=
a_{\ell m}^{\scriptscriptstyle\rm EE}$ for $m=0,\pm6,\pm12,\pm18,\cdots$ and 
$a_{\ell m}^{\scriptscriptstyle\rm AA}=
-a_{\ell m}^{\scriptscriptstyle\rm EE}$ for $m=\pm2,\pm4,\pm 8,\cdots$. 
Thus, one finds
\begin{equation}
\sigma_{\ell}^{\scriptscriptstyle\rm  AA}
=\sigma_{\ell}^{\scriptscriptstyle\rm EE}.
\label{eq:identity_AA_EE}
\end{equation}

The similar identity also holds for the cross-correlated signals 
$\mathcal{F}_{\scriptscriptstyle\rm AT}$ and $\mathcal{F}_{\scriptscriptstyle\rm ET}$. 
Applying the relation (\ref{eq:Rotation a_{em}-self-cross}) to the expressions (\ref{eq:Antenna, AE,AT,ET}), one obtains 
\begin{eqnarray}
a_{\ell m}^{\scriptscriptstyle\rm ET} &=& 
-\frac{2\sqrt{2}}{3}\,e^{-i\,m\delta}\,
\sin^2\left(\frac{m\delta}{2}\right)
\left\{ a_{\ell m}^{\scriptscriptstyle\rm S_1S_1}+ 
a_{\ell m}^{\scriptscriptstyle\rm S_1S_2} + 
(-1)^m\,[a_{\ell m}^{\scriptscriptstyle\rm S_1S_2}]^* \right\}
\nonumber\\
&=& -\frac{i}{\sqrt{3}}\,\tan\left(\frac{m\delta}{2}\right)\,\,
a_{\ell m}^{\scriptscriptstyle\rm AT}.
\label{eq:a_em ET, AT-Sij}
\end{eqnarray}
Thus, both of the multipole moments $a_{\ell m}^{\scriptscriptstyle\rm AT}$ and $a_{\ell m}^{\scriptscriptstyle\rm ET}$ become vanishing when $m=0,\,\pm3,\,\pm6,\cdots$. Further, for all the non-vanishing components, the absolute value of the pre-factor becomes unity. This immediately yields the relation:   
\begin{equation}
\sigma_{\ell}^{\scriptscriptstyle\rm AT}= 
\sigma_{\ell}^{\scriptscriptstyle\rm ET}.
\label{eq:{eq:identity_AT_ET}}
\end{equation}
Note that while the relations (\ref{eq:identity_AA_EE}) and 
(\ref{eq:{eq:identity_AT_ET}}) are derived in a specific choice of the coordinate system (\ref{eq: a,b,c}), the final results do not depend on the coordinates.

Finally, we note a quite remarkable fact for the cross-correlated signals, $\mathcal{F}_{\scriptscriptstyle\rm AE}$, 
$\mathcal{F}_{\scriptscriptstyle\rm AT}$ and 
$\mathcal{F}_{\scriptscriptstyle\rm ET}$. 
It is shown in Appendix \ref{appendix:proof} that the multipole moments 
$\ell=0$ and $1$ for the antenna pattern 
$\mathcal{F}_{\scriptscriptstyle\rm AE}$ are 
{\it exactly} zero, while the monopole mode ($\ell=0$) vanishes for the cross-correlated signals  
$\mathcal{F}_{\scriptscriptstyle\rm AT}$ and 
$\mathcal{F}_{\scriptscriptstyle\rm ET}$, over the whole frequency range: 
\begin{eqnarray}
&&   \sigma_0 (f) =0,\quad  \sigma_1 (f) =0 ~~~\text{for AE-correlation},
\cr
&&   \sigma_0 (f) =0 \quad \quad \quad \quad ~~~~~~~~
\text{for AT-,ET-correlation}.
\label{eq:sigma^AE=0 for ell=0,1}
\end{eqnarray}
Here, the important properties of the antenna pattern functions derived from the parity invariance and geometric argument are summarized in 
Table \ref{tab:summay_multipole}.

\begin{table*}
\caption{\label{tab:summay_multipole}
Symmetric properties of antenna pattern function}
\begin{ruledtabular}
\begin{tabular}{ccl}
Combination of variables & Condition & Properties of $\sigma_{\ell}(f)$ \\
\hline
All & low-frequency limit $(\hat{f}\ll1)$ & 
$\sigma_{\ell}=0$ for $\ell\ne0,~2,~4$     \\
All & self-correlation &$\sigma_{\ell}=0$  for $^{\forall}\ell=$odd
\\
(A,A), (E,E) & & $\sigma_{\ell}^{\scriptscriptstyle\rm AA}
=\sigma_{\ell}^{\scriptscriptstyle\rm EE}$ 
\\
(A,E) & &$\sigma_{\ell}=0$ for $\ell=0,~1$$^{\dagger}$ 
\\
(A,T), (E,T) & & $\sigma_{\ell}=0$ for $\ell=0$$^{\dagger}$, 
$\sigma_{\ell}^{\scriptscriptstyle\rm AT}=
\sigma_{\ell}^{\scriptscriptstyle\rm ET}$
\\
\end{tabular}
\end{ruledtabular}
\leftline{$^{\dagger}$ The details of the proof are presented in 
Appendix.\ref{appendix:proof}}
\end{table*}

\section{Angular power and directional sensitivity of space interferometer}
\label{sec:directional sensitivity}

While several important properties for directional sensitivity of the 
space interferometer were found in previous section, 
it remains  still unclear how the multipole moments of the antenna 
pattern functions quantitatively depend on the frequency and the 
combinations of data streams. In this section, based on the previous 
remarks, the spherical harmonic analysis of the antenna pattern function 
is carried out analytically and numerically in specific choices of the 
data combinations.  
For a relevant range of the frequencies beyond the low-frequency 
approximation, the directional sensitivity to the antenna pattern is 
estimated in the LISA case, taking into account the instrumental noises.

\subsection{Toy model example}
\label{subsec:a toy model}

As noted in Sec.\ref{sec:Parity of detector response}, the frequency and 
angular dependences of the transfer function $\mathcal{T}$ in equation 
(\ref{eq:one-arm_response}) make it difficult to treat the spherical 
harmonic analysis of the antenna pattern. 
If we set $\mathcal{T}=1$, however, the spherical harmonic expansion of 
the antenna pattern can be analytically evaluated, the results of which 
are compared with the realistic cases without invoking the assumption 
$\mathcal{T}=1$.

\begin{figure}[ht]
\begin{center}
\includegraphics[width=7cm,clip]{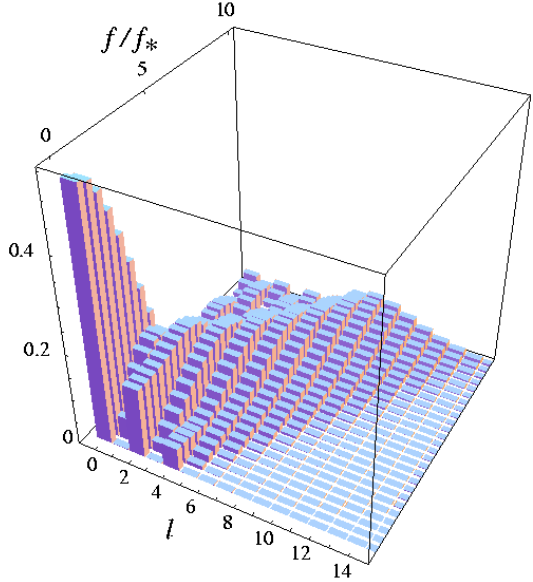}
\caption{
Angular power $\sigma_{\ell}(\hat{f})$ of the antenna pattern function 
for the toy model as a function of $\ell$ and $\hat{f}$.
\label{fig:1}
}
\end{center}
\end{figure}

For computational purpose in this subsection, we set the directional 
unit vectors for three spacecrafts as: 
\begin{eqnarray}
{\bf{a}} = {\bf{z}}, 
\quad
{\bf{b}} =  \frac{\sqrt{3}}{2} {\bf{y}} - \frac{1}{2} {\bf{z}}, 
\quad
{\bf{c}} = - \frac{\sqrt{3}}{2} {\bf{y}} - \frac{1}{2} {\bf{z}}, 
\label{eq: a,b,c_toy_model}
\end{eqnarray}
and consider the antenna pattern for Michelson signal. 
Table \ref{tab:summay_multipole} suggests that 
a number of non-vanishing multipole moments is severely 
restricted in the case of the self-correlated signals, since the 
assumption $\mathcal{T}=1$ roughly corresponds to the low-frequency limit. 
An interesting case is therefore to take the cross-correlation 
between the signals extracted from the vertices $1$ and $2$. 
In this case, the response function at the rest frame becomes
\begin{eqnarray}
     \mathcal{F}_{12} (f, {\bf\Omega}) 
     &=& 
      e^{ i \hat{f}\, {\bf\Omega}{\cdot} 
            \textbf{a} }
        \sum_{A=+,\times} 
        F^{A}_1(  {\bf\Omega},f) F^{A}_2(  {\bf\Omega},f)  
        ~~~;  ~~~~
\left\{
\begin{array}{c}
F^{A}_1 = \frac{1}{2}[{\bf a}\otimes{\bf a}
 -{\bf c}\otimes{\bf c}]:{\bf e}^A({\bf\Omega}) 
\\
\\
F^{A}_2 = \frac{1}{2}[{\bf b}\otimes{\bf b}
 -{\bf a}\otimes{\bf a}]:{\bf e}^A({\bf\Omega}) 
\end{array}.
\right.\nonumber
\end{eqnarray}
With the specific choice of the coordinate system 
(\ref{eq: a,b,c_toy_model}), the explicit expression for the antenna 
pattern becomes 
\begin{eqnarray}
\mathcal{F}_{12} (f, {\bf\Omega}) 
&=& -\frac{3}{64}\left[3\cos^4\phi+ 2\sin^2\theta\cos^2\phi +
                  3\sin^4\theta \right.
\nonumber \\
&&~~~~~~~~~~~~~~\left.+\cos^2\theta\sin^2\phi\left\{
-2+5\cos2\theta+2\cos2\phi+3\cos^2\theta \sin^2\phi\right\}\right]
e^{ - i \hat{f} \cos\theta }.
\label{eq:toy model-R_cc of}
\end{eqnarray}
Note that the function (\ref{eq:toy model-R_cc of}) possesses the following symmetry: 
\begin{alignat}{4}
         \mathbb{Q\,P}\,\mathcal{F}_{12}
         & =  \left[ \mathcal{F}_{12}\right]^*, \quad &
         \mathbb{P}\,\mathcal{F}_{12} 
         & =   \mathcal{F}_{12} , \quad &
         \mathbb{Q}\,\mathcal{F}_{12} 
         & =  \left[ \mathcal{F}_{12} \right]^*  \quad &.  
\label{eq:PQ rule for a toy model}
\end{alignat}
This indicates that the antenna pattern is sensitive to both the even 
and the odd modes, while the multipole moments $a_{\ell m}$ with 
$m=\mathrm{odd}$ become vanishing. 
Since the relation (\ref{eq:parity relation of a_l-m}) always holds,  
it is sufficient to treat the case for $m\geq0$ only. 
Substituting the explicit expression (\ref{eq:toy model-R_cc of}) into 
the definition (\ref{eq:a_lm_in_detector_frame}), the integral over 
$\phi$ is first performed. Writing $\cos\theta$ by $u$, we have 
\begin{eqnarray}
    a_{\ell m}(\hat{f})
    = \sqrt{ \frac{(2\ell+1)}{4\pi} 
             \frac{(\ell-m)!}{(\ell+m)!}}
    \int_{-1}^{1} du\,e^{-i\hat{f}u} P_{\ell}^m(u)\,g_m(u),
\label{eq:a_lm_reduced}
\end{eqnarray}
where the function $g_m(u)$ can be expressed as polynomial series as 
\begin{equation}
    g_m(u)=\sum_{N=0}^2 c^{(m,N)} u^{2N}(1-u^2) ^{|m|/2}.
\label{eq:func_gm}
\end{equation}
The coefficients $ c^{(m,N)}$ are the numerical constants, which are 
summarized in Appendix \ref{appendix:toy model}. 
Note that the function $g_m$ are non-vanishing only for 
$m=0,\,\pm2,\,\,\pm4$, indicating that the non-vanishing components of 
$a_{\ell m}$ are obtained only when $m=0,\,\pm2,\,\,\pm4$. 
From (\ref{eq:a_lm_reduced}) and (\ref{eq:func_gm}), the remaining 
integrals become of the form:
\begin{equation}
    {\cal I}_{\ell m}^{N} 
    = \int_{-1}^{1}du\,\, e^{-i\hat{f}u} u^{2N} 
       (1-u^2)^{|m|/2}\,P_{\ell}^m(u).
\end{equation}
This integral is analytically performed according to 
Ref. \cite{Allen:1997gp}. Using the formula for  
Legendre polynomials,  
$P_{\ell}^m(u)=(-1)^m(1-u^2)^{m/2}(d/du)^mP_{\ell}(u)$ for $m\geq0$, 
repeating the integration by parts yields   
\begin{eqnarray}
  {\cal I}_{\ell m}^{N}
     &=&  \int_{-1}^{1}\,\, du\,\,P_{\ell}(u)\, 
          \frac{d^m}{du^m}\left[e^{-i\hat{f}\,u}\,u^{2N} (1-u^2)^{m}\right]
    \nonumber
    \\
    &=& \sum_{s=0}^{2(m+N)}\,g_s^{(m,N)}(\hat{f})\,\,
    \int_{-1}^{1} du P_{\ell}(u)\,u^s\,e^{-i\hat{f}u}.
    \label{eq:cal_I}
\end{eqnarray}
The quantities $g_s^{(m,N)}(\hat{f})$ are the polynomial function of
$\hat{f}$ up to the forth order and are listed in Appendix
\ref{appendix:toy model}. 
The integral in the last line is expressed in terms of the spherical
Bessel function $j_{\ell}$:  
\begin{eqnarray}
    \int_{-1}^{1} du P_{\ell}(u) u^s  e^{-i\hat{f}u} 
    &=& i^s \frac{d^s}{d\hat{f}^s} \int_{-1}^{1}du ~
       P_{\ell}(u) e^{- i\hat{f} u}
    \nonumber
    \\
    &=& 2 (-1)^\ell \,\,i^{\ell+s} \frac{d^s}{d\hat{f}^s}
        \,\, j_{\ell}(\hat{f}).
\label{eq:toy model-formula_besselj}
\end{eqnarray}
Thus, substituting the results (\ref{eq:func_gm}), (\ref{eq:cal_I}) 
and (\ref{eq:toy model-formula_besselj}) into (\ref{eq:a_lm_reduced}), 
one finally obtains the analytic expression for multipole coefficients: 
\begin{eqnarray}
   a_{\ell m}(\hat{f}) 
   = 
   \sqrt{\frac{(2\ell+1)}{\pi} \frac{(\ell-m)!}{(\ell+m)!}}  
   \sum_{N=0}^{2} \sum_{s=0}^{2(m+N)}(-1)^\ell \,\,i^{\ell+s}\,\,   
   c^{(m,N)} g_s^{(m,N)}(\hat{f}) \,  
   \frac{d^s}{d \hat{f}^s} j_{\ell} (\hat{f}), \qquad (m\geq0). 
\label{eq:a_lm_toy_model} 
\end{eqnarray}
While the above expression is the outcome based on the 
coordinate (\ref{eq: a,b,c_toy_model}), the invariant quantity 
$\sigma_{\ell}(\hat{f})$ can be evaluated from 
(\ref{eq:a_lm_toy_model}), which is depicted  in 
Figure \ref{fig:1} as a function of $\ell$ and $\hat{f}$. Also using 
(\ref{eq:a_lm_toy_model}), the non-vanishing components of 
$\sigma_{\ell}$ in the low-frequency limit are explicitly calculated as 
\begin{eqnarray}
&&\sigma_0=\frac{3\sqrt{\pi}}{10} - \frac{\sqrt{\pi}}{28}\hat{f}^2, 
\quad
\sigma_1=\frac{\sqrt{2\pi}}{14}\hat{f} - 
\frac{17\sqrt{2\pi}}{2520}\hat{f}^3, 
\quad
\sigma_2=\frac{3\sqrt{\pi}}{35} + \frac{(\sqrt{30}-6)\sqrt{\pi}}{960}
\hat{f}^2, 
\nonumber\\
&&\sigma_3=\frac{1}{42}\sqrt{\frac{19\pi}{10}}\hat{f} 
- \frac{43}{1848}\sqrt{\frac{\pi}{190}}\hat{f}^3, 
\quad
\sigma_4=\frac{1}{140}\sqrt{\frac{47\pi}{3}} - 
\frac{19}{264}\sqrt{\frac{19\pi}{141}}\hat{f}^2
\label{eq:sigma_toy_model}
\end{eqnarray}
up to the third order in $\hat{f}$. The leading order terms in 
$\sigma_{\ell}$ rigorously match the results in equation 
(\ref{eq:sigma_cross_michelson_low_freq}).

Figure \ref{fig:1} shows that the higher multipole moments appear as 
increasing the frequency, and an oscillatory behavior is found in the 
frequency domain $\hat{f}\gtrsim1$, 
which are also indicated from the low-frequency expansion in equation 
(\ref{eq:sigma_toy_model}). From the analytic expression 
(\ref{eq:a_lm_toy_model}), we readily see that the quantity 
$\sigma_{\ell}$ higher than $\ell\gtrsim4$ scale as 
$\mathcal{O}(\hat{f}^{\ell-4})$ in the low-frequency limit and 
asymptotically behaves as $\sigma_{\ell}\propto\hat{f}^{-1}$ in 
the high-frequency limit. The resultant angular power depicted in 
Figure \ref{fig:1} implies that the resolution of anisotropy in the 
stochastic background of gravitational waves can reach 
$\ell\lesssim10 \sim 15$ for a relevant frequency range  
$0\leq\hat{f}\lesssim10$. This result is comparable to the angular 
resolution of gravitational-wave background measured from the 
ground-based detectors~\cite{Allen:1997gp,Cornish:2001hg}, since 
the assumption neglecting the frequency and the directional dependences 
of transfer function $\mathcal{T}$ can be validated for the response 
function of Fabry-Perot interferometer.

\subsection{Influence of transfer function}

We now calculate the angular power of the antenna pattern fully taking 
into account the frequency and the angular dependences of the transfer 
function $\mathcal{T}$. The results are then compared with the toy model 
example. For this purpose, the spherical harmonic expansion of the 
antenna pattern function is numerically carried out using the 
{\it SPHEREPACK} 3.1 package \citep{Adams:2003}.

\begin{center}
\begin{figure}[ht]
\begin{center}
\includegraphics[width=7cm,clip]{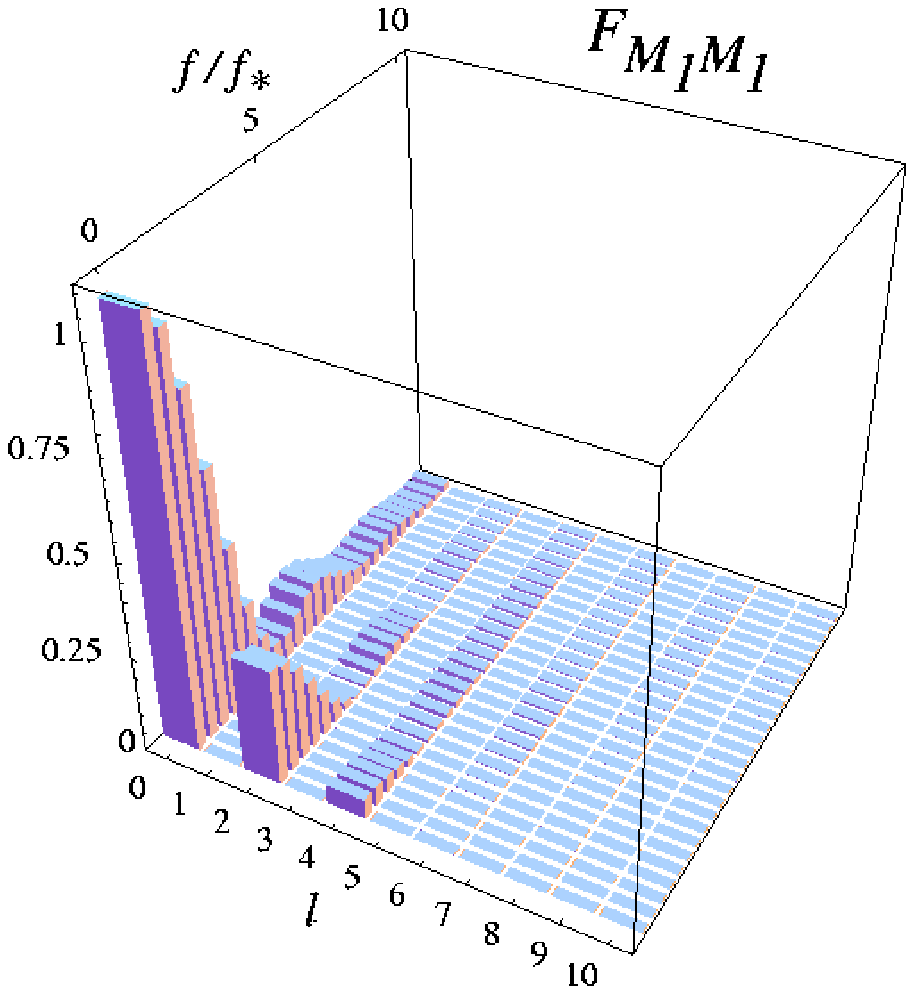}
\hspace{0.5cm}
\includegraphics[width=7cm,clip]{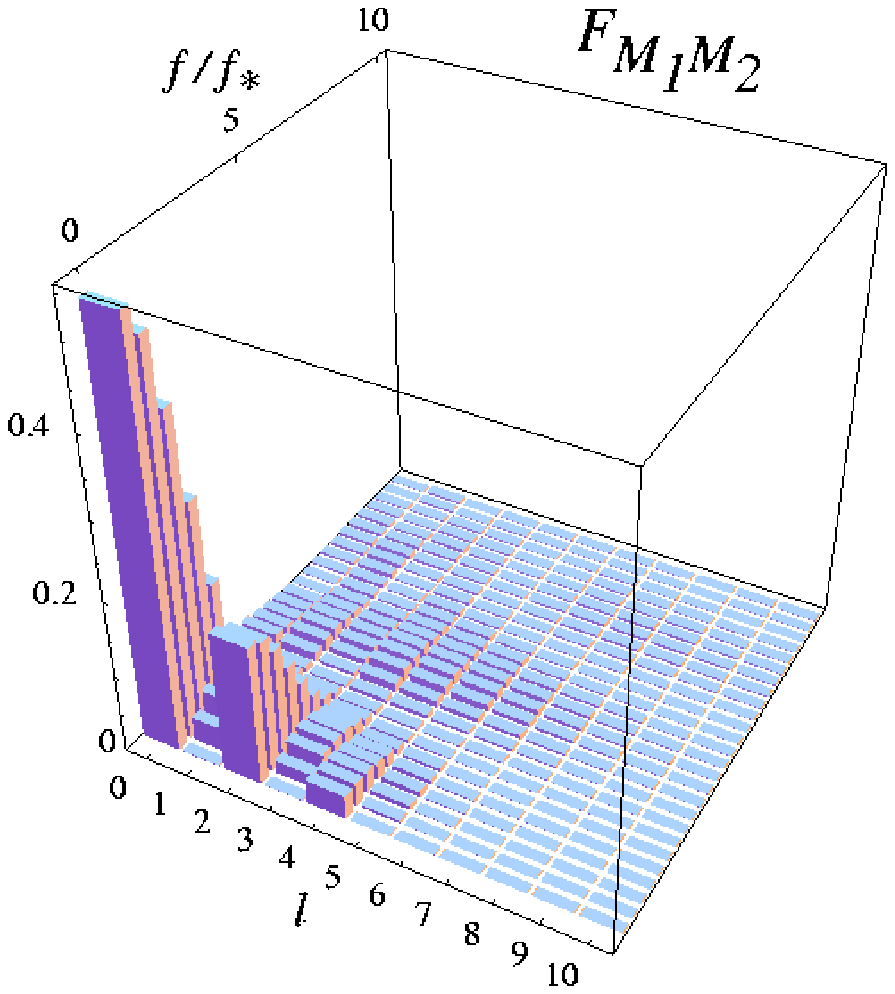}
\end{center}
\caption{
Angular power $\sigma_{\ell}(\hat{f})$ of the antenna pattern function 
for the self-correlated Michelson signals 
$\mathcal{F}_{\scriptscriptstyle\rm M_1M_1}$({\it left}) 
and the cross-correlated 
Michelson signals, $\mathcal{F}_{\scriptscriptstyle\rm M_1M_2}$ 
({\it right}). 
\label{fig:sigma_michelson}
}
\end{figure}
\end{center}

Figure \ref{fig:sigma_michelson} shows the angular power of the antenna 
pattern function for the self-correlated 
Michelson signals $\mathcal{F}_{\scriptscriptstyle\rm M_1M_1}$({\it
left}) 
and the cross-correlated Michelson signals 
$\mathcal{F}_{\scriptscriptstyle\rm M_1M_2}$({\it right}). 
Relaxation of the assumption 
$\mathcal{T}=\mbox{const.}$, i.e., low-frequency approximation, leads to 
the non-vanishing components for even modes with $\ell\geq6$. However, 
the resultant higher multipole moments turn out to be highly
suppressed. While the sensitivity to the higher multipole moments is 
slightly improved in the case of the cross-correlated Michelson signals, 
comparing it with Fig. \ref{fig:1} reveals that the frequency 
dependence of the transfer function $\mathcal{T}$ significantly 
reduces the angular power in both the lower and the higher multipole 
moments. The numerical evaluation of spherical harmonic expansion 
implies that the non-vanishing components of the angular power 
asymptotically decrease as $\sigma_{\ell}\propto
\hat{f}^{-2}$ in the high frequency region, even faster than that 
of the toy model example.

\begin{center}
\begin{figure}[ht]
\begin{center}
\includegraphics[width=7cm,clip]{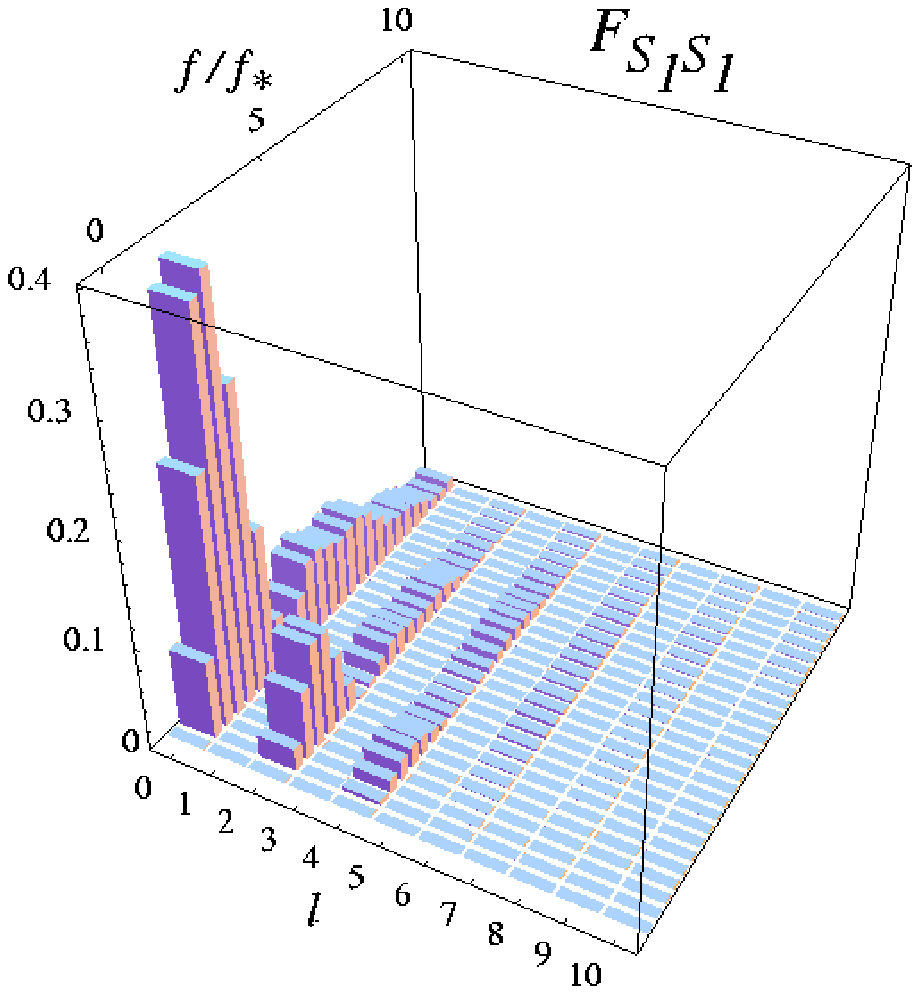}
\hspace{0.5cm}
\includegraphics[width=7cm,clip]{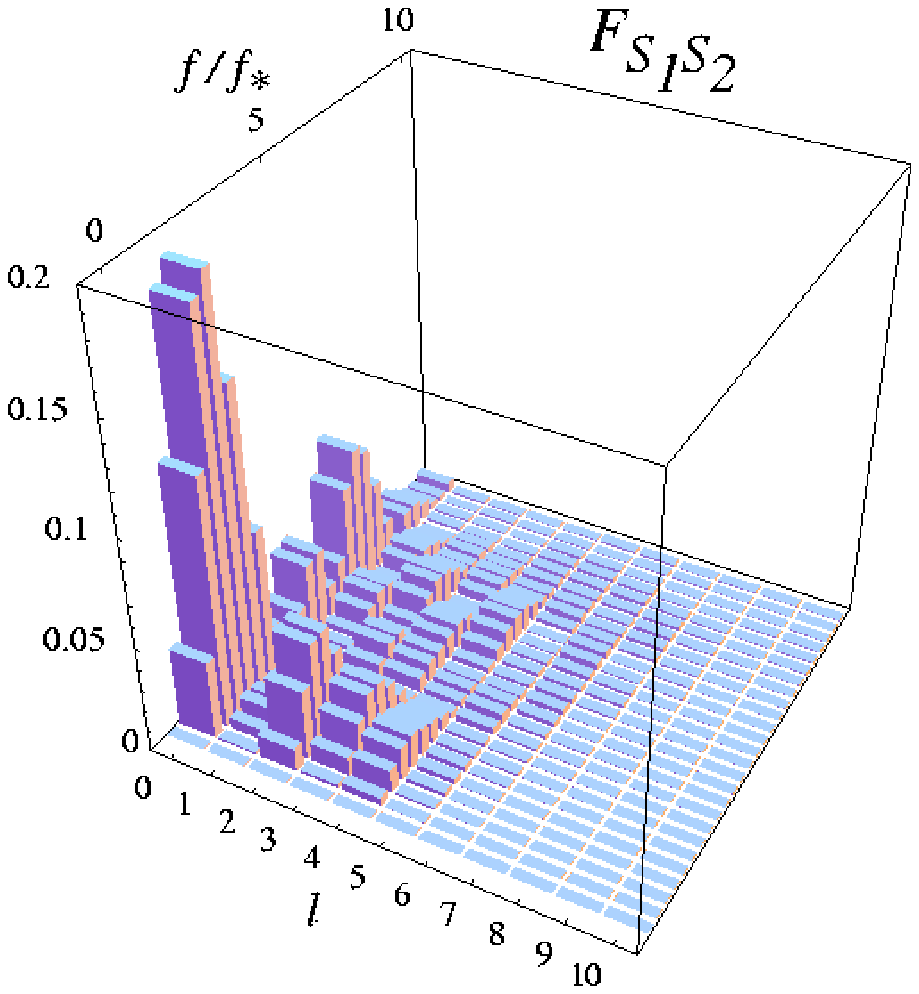}
\end{center}
\caption{
Angular power $\sigma_{\ell}(\hat{f})$ of the antenna pattern function 
for the self-correlated Sagnac signals 
$\mathcal{F}_{\scriptscriptstyle\rm S_1S_1}$
({\it left}) and the cross-correlated 
Sagnac signals $\mathcal{F}_{\scriptscriptstyle\rm S_1S_2}$({\it right}) . 
\label{fig:sigma_sagnac}
}
\end{figure}
\end{center}

The behaviors of the angular power are qualitatively similar to the 
case adopting the Sagnac variables that cancel the laser frequency 
noise (Fig.\ref{fig:sigma_sagnac}). Apart from the low-frequency 
limit, where the antenna pattern function for Sagnac signals behaves as 
$\mathcal{F}_{\scriptscriptstyle {\rm S}_i{\rm S}_j}\sim \hat{f}^2$ 
[see Eqs.~(\ref{eq:antenna_self_correlation_low_freq}) and 
(\ref{eq:antenna_cross_correlation_low_freq})], the angular power is 
highly suppressed at the frequency $\hat{f}\gtrsim1$ even in the 
relatively lower multipole moments $\ell\lesssim6$. Thus, the 
directional sensitivity of the space interferometer to a stochastic 
background of gravitational waves is severely limited by the frequency 
dependence of the transfer function. 
This fact is irrelevant to the choice of the interferometric variables.

%
%
%

\subsection{Directional sensitivity for optimal combinations of TDI variables}
\label{subsec:sensitivity}

To elucidate a more quantitative aspect of the directional sensitivity 
to the gravitational-wave background, it will need to take into 
account effects of detector noises. 
To investigate this, rather than using the Michelson and Sagnac 
signals, a set of optimal TDIs (A, E, T) free from the noise 
correlations should be applied to the correlation analysis of the 
gravitational-wave signals.

\begin{center}
\begin{figure}[ht]
\begin{center}
\includegraphics[width=7cm,clip]{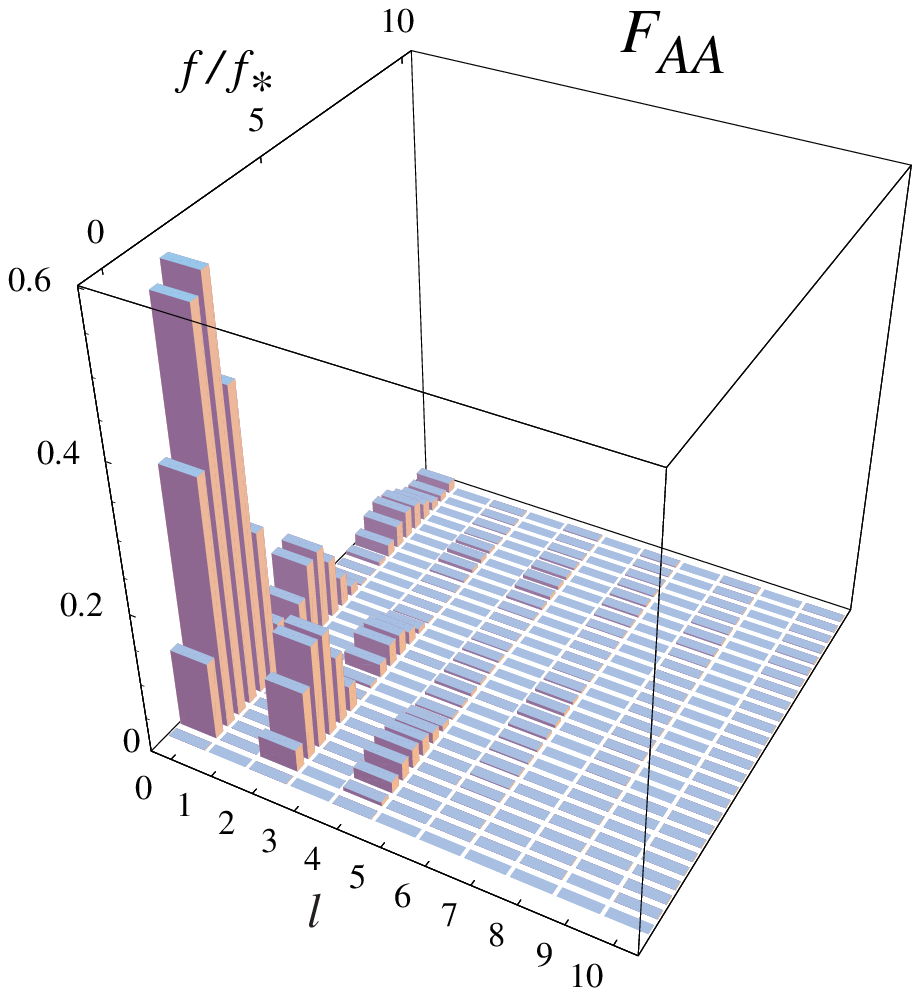}
\hspace{0.5cm}
\includegraphics[width=7cm,clip]{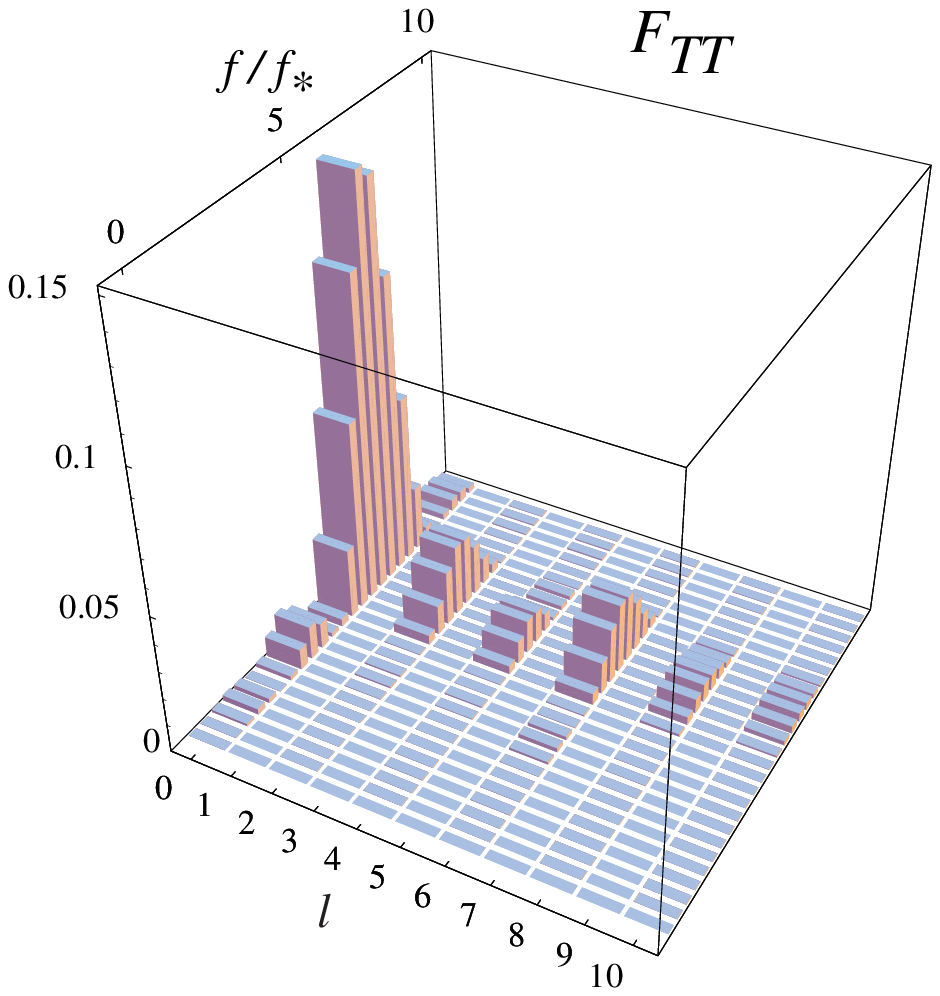}
\end{center}
\begin{center}
\includegraphics[width=7cm,clip]{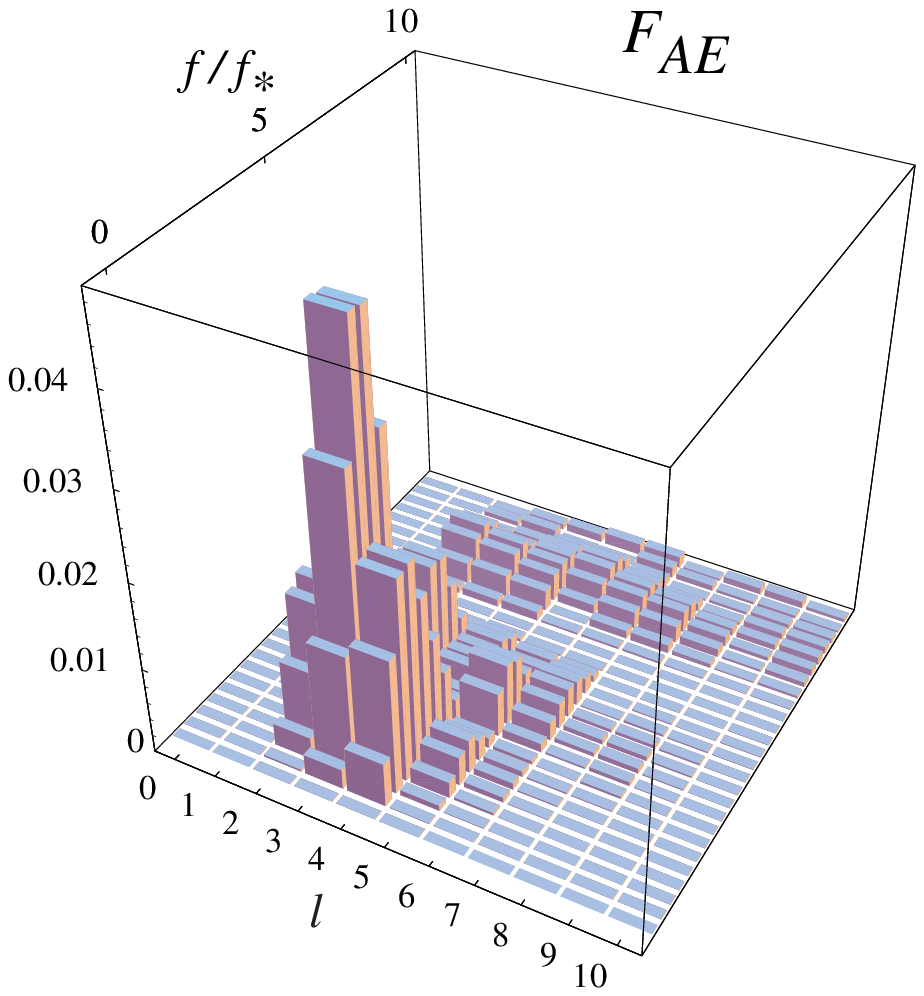}
\hspace{0.5cm}
\includegraphics[width=7cm,clip]{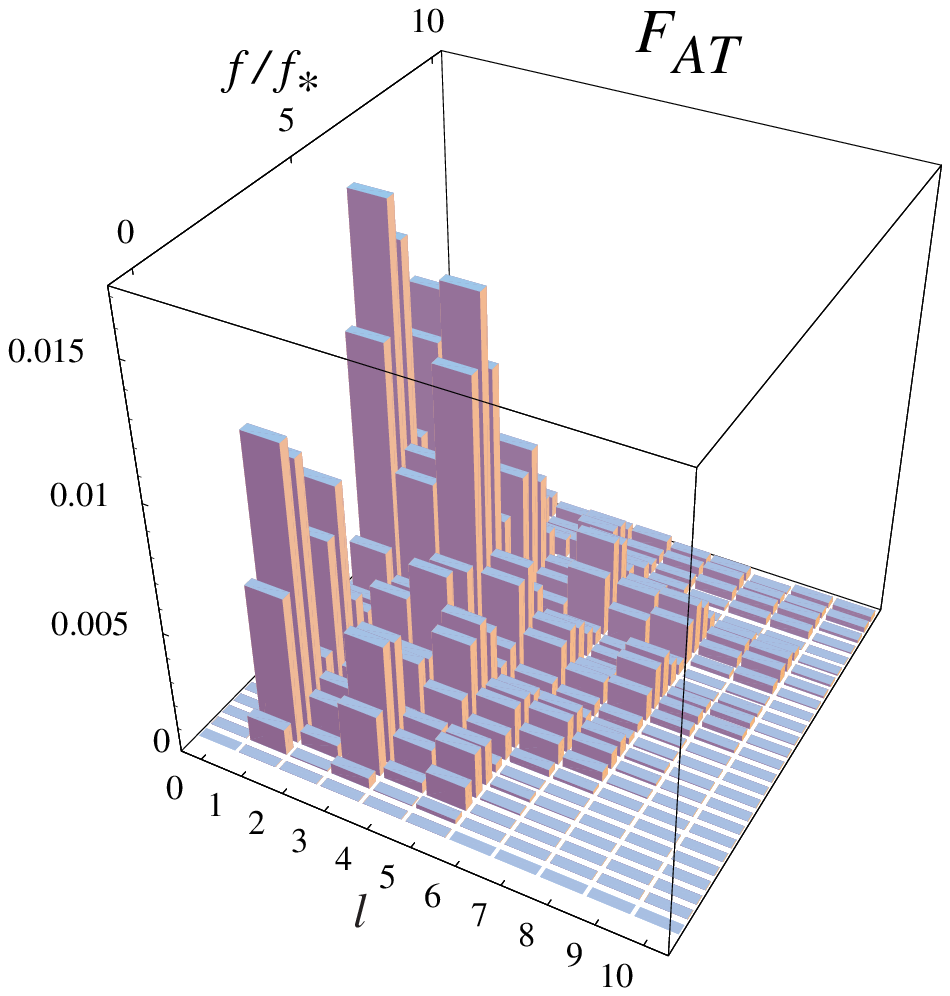}
\end{center}
\caption{
Angular power $\sigma_\ell (\hat{f})$ of the antenna pattern 
functions for the optimal TDI variables. The top panels show the 
magnitude $\sigma_{\ell}$ in the self-correlated cases 
$\mathcal{F}_{\scriptscriptstyle\rm AA}$({\it left}) and 
$\mathcal{F}_{\scriptscriptstyle\rm TT}$({\it right}), 
while the bottom panels represent 
the results obtained from the cross-correlated signals 
$\mathcal{F}_{\scriptscriptstyle\rm AE}$({\it left}) and 
$\mathcal{F}_{\scriptscriptstyle\rm AT}$({\it right}). 
Note that the angular powers of antenna pattern 
$\mathcal{F}_{\scriptscriptstyle\rm EE}$ and 
$\mathcal{F}_{\scriptscriptstyle\rm ET}$ 
coincide with those of $\mathcal{F}_{\scriptscriptstyle\rm AA}$ and 
$\mathcal{F}_{\scriptscriptstyle\rm AT}$. 
\label{fig:sigma_optimal_TDIs}
}
\end{figure}
\end{center}

Figure \ref{fig:sigma_optimal_TDIs} plots the quantity 
$\sigma_{\ell}$ for various combinations of the optimal TDIs. 
Note that the angular power of the antenna pattern function 
$\mathcal{F}_{\scriptscriptstyle\rm EE}$ 
($\mathcal{F}_{\scriptscriptstyle\rm ET}$) coincides with that 
obtained from 
$\mathcal{F}_{\scriptscriptstyle\rm AA}$
($\mathcal{F}_{\scriptscriptstyle\rm  AT}$), although the sky patterns 
themselves differ from each other. 
For the self-correlated signals, the amplitude $\sigma_{\ell}$ of 
$\mathcal{F}_{\scriptscriptstyle\rm AA}$ is quite 
similar to that of the self-correlated 
Sagnac signals $\mathcal{F}_{\scriptscriptstyle\rm  S_1S_1}$, 
while the low-frequency part 
of the angular power for $\mathcal{F}_{\scriptscriptstyle\rm  TT}$ 
is highly suppressed, 
which can be deduced from the low-frequency approximation presented 
in Appendix \ref{appendix:1}. 
As for the cross-correlation signals, the monopole and the dipole 
moments for the antenna pattern function
$\mathcal{F}_{\scriptscriptstyle\rm  AE}$ are 
exactly canceled and the monopole moment for
$\mathcal{F}_{\scriptscriptstyle\rm  AT}$ 
further vanishes (Table \ref{tab:summay_multipole} and Appendix 
\ref{appendix:proof}). 
Apart from these facts, the magnitude $\sigma_{\ell}$ at frequency 
$\hat{f}\gtrsim1$ shows a rich structure with many peaks, indicating 
that the directional sensitivity could be improved at $\hat{f}\gtrsim1$. 
As shown in Fig. \ref{fig:sigma_optimal_TDIs}, the angular power of 
the cross-correlated signals is one order of magnitude smaller than 
that of the self-correlated signals, however, this does not directly 
imply that the self-correlated signals are more sensitive to an 
anisotropy of a gravitational-wave background. 

Based on these results, let us now quantify the directional 
sensitivity of the antenna pattern function. 
Assuming that the laser frequency noise can be either canceled or 
sufficiently reduced by the TDI technique with the use of the recently 
proposed laser self-locking
scheme~\cite{Sheard:2003fz,Sylvestre:2004pm}, the dominant noise
contributions to detector's output would be the 
acceleration noise and the shot noise. According to 
Ref.~\cite{Cornish:2001bb}, the noise spectral densities for optimal 
TDIs are calculated as (see also \cite{Prince:2002hp}):
\begin{eqnarray}
S_n^{\scriptscriptstyle\rm AA}(f) &=& S_n^{\scriptscriptstyle\rm EE}(f) 
= \sin^2(\hat{f}/2)\,\,\left\{8\,\left(2+ \cos\hat{f}\right)\,\,
S_{\rm shot}(f) + 16\,\left(3 +2 \cos\hat{f}+\cos2\hat{f} \right)\,\,
S_{\rm accel}(f)\right\},
\cr
S_n^{\scriptscriptstyle\rm TT}(f)&=& 2\,\,
\left(1+2\cos\hat{f}\right)^2\left\{ S_{\rm shot}(f) + 
4\sin^2(\hat{f}/2)\,\,
S_{\rm accel}(f) \right\}. 
\label{eq:noise_TT}
\end{eqnarray}
Note that the cross-correlated noise spectra are exactly canceled. 
Here we specifically adopt the noise functions for the LISA detector:
$S_{\rm shot}(f)=4.84\time10^{-42}$Hz$^{-1}$ and 
$S_{\rm accel}(f)=2.31\time10^{-40}(\mbox{mHz}/f)^4$Hz$^{-1}
$\cite{Cornish:2001bb}. We then define the {\it effective sensitivity} 
for the multipole moment $\ell$, $h_{\rm eff}^{(\ell)}(f)$, which 
characterizes the rms amplitude of the noise-to-angular power ratio:  
\begin{eqnarray}
    h^{(\ell)}_{\rm eff} (f) 
= (4\pi)^{1/4}\,\,\sqrt{\frac{S_{\rm n}(f)}{\sigma_\ell (f)}} 
\label{eq:h_eff_l_self}
\end{eqnarray}
for self-correlation signals. Setting $\ell=0$, the above definition 
recovers the usual meaning of sensitivity curve. Thus, the quantity 
$h_{\rm eff}^{(\ell)}(f)$ may be regarded as the effective power 
of $\ell$-th moment relative to the monopole moment as a reference 
sensitivity. 
For the cross-correlated signals, on the other hand, the absence of 
noise correlation implies that the signal-to-noise ratio can be 
improved by optimally filtering the cross-correlated signals. 
The resultant form of the signal-to-noise ratio shows the explicit 
dependence on the observation time $T$ \cite{Allen:1997ad}. 
According to Ref.~\cite{Cornish:2001bb}, the effective sensitivity 
for cross-correlated signals may be written as: 
\begin{eqnarray}  
    h_{\rm eff}^{(\ell)} (f) 
    = \left(\frac{4\pi}{T\Delta f}\right)^{1/4}\, 
    \left[ \frac{S_{n,1}(f)\,\,S_{n,2}(f)}
        {\sigma^2_{\ell}(\hat{f})} \right]^{1/4},
\label{eq:h_eff_l_cross}
\end{eqnarray}  
where $\Delta f$ denotes the frequency resolution for actual output data.

\begin{center}
\begin{figure}[ht]
\begin{tabular}{ll}
\begin{minipage}{70mm}
\begin{center}
\includegraphics[width=7cm,clip]{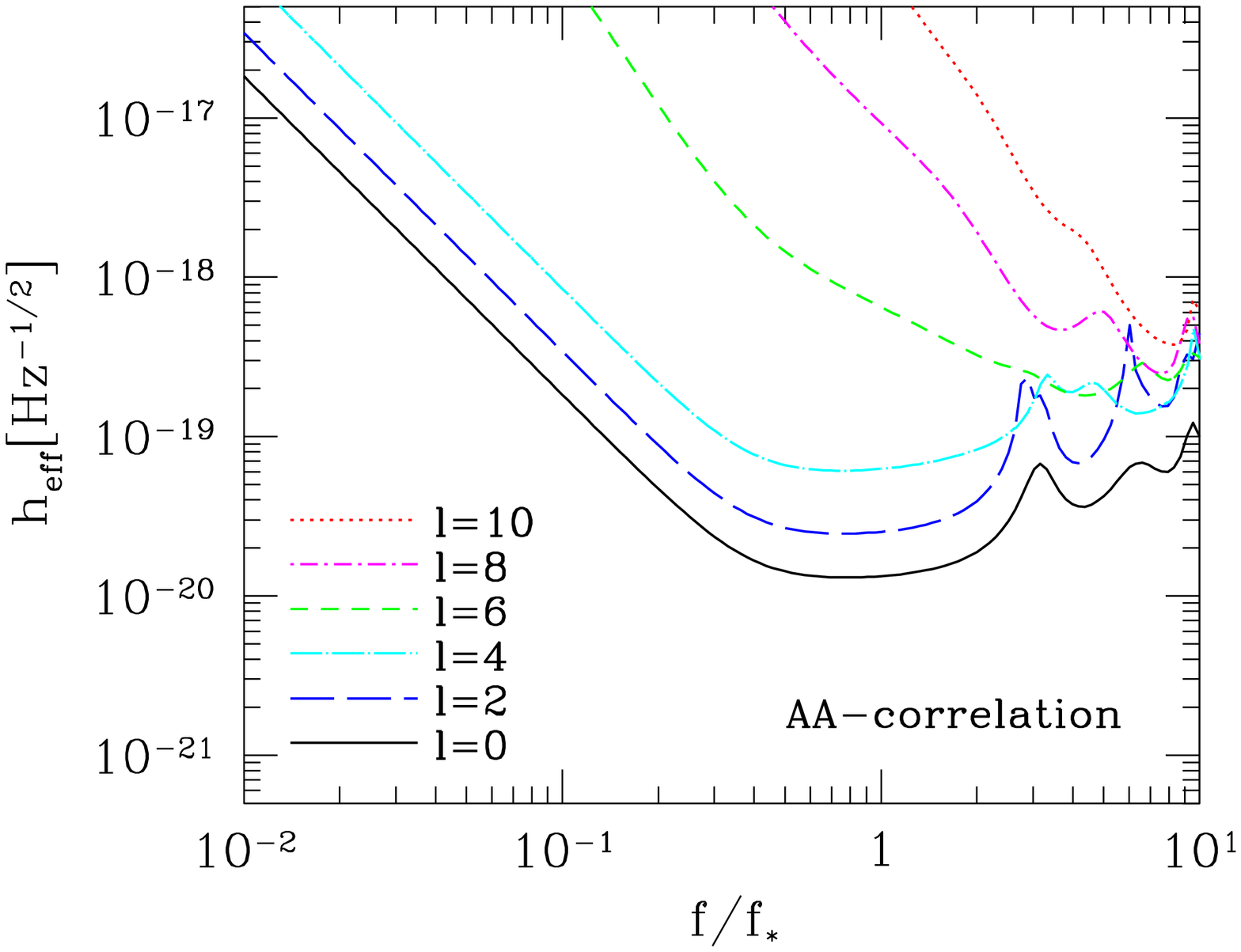}
\end{center}
\end{minipage}
&\hspace{0.5cm}
\begin{minipage}{70mm}
\begin{center}
\includegraphics[width=7cm,clip]{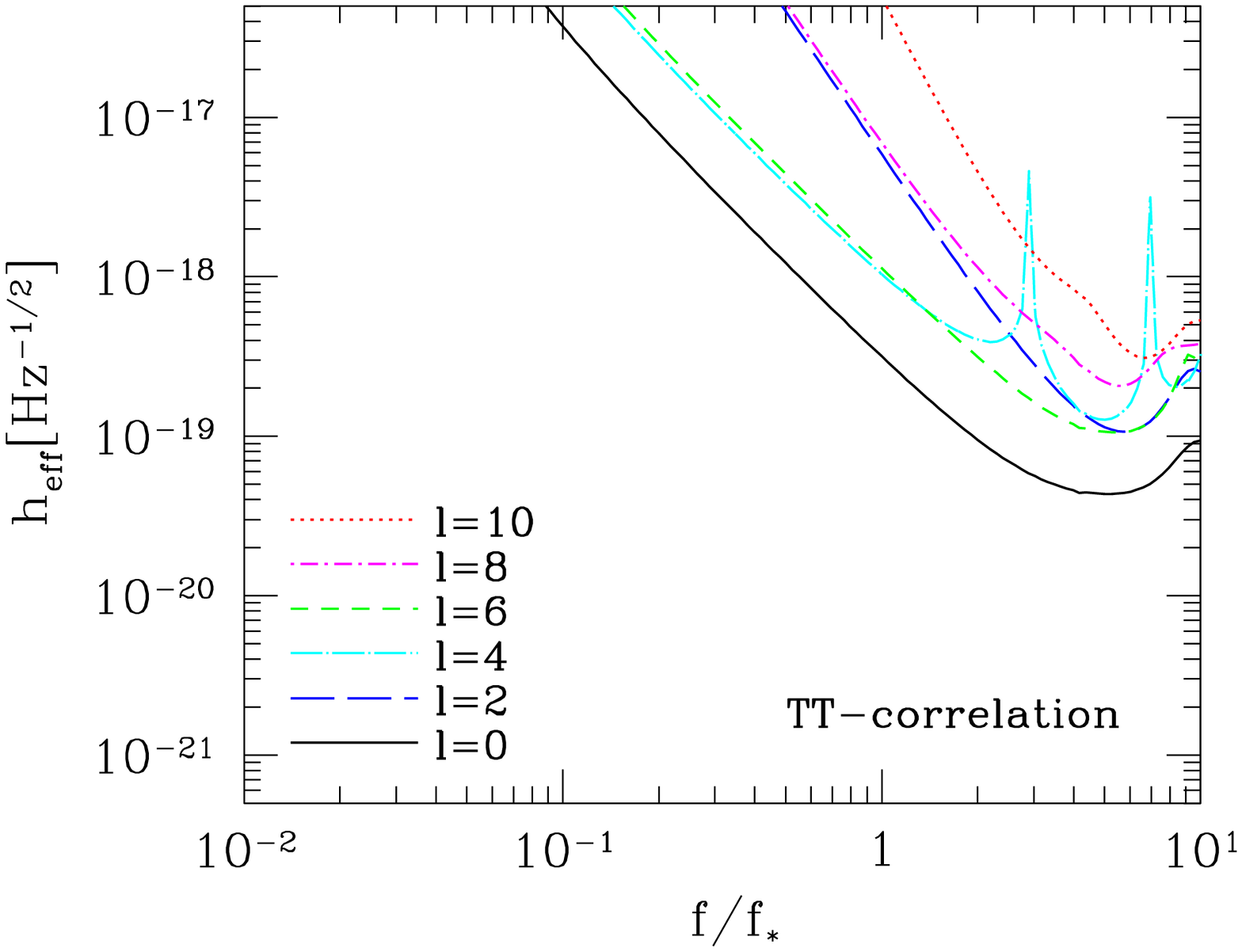}
\end{center}
\end{minipage}
\\[5mm]
\begin{minipage}{70mm}
\vspace{5mm}
\begin{center}
\includegraphics[width=7cm,clip]{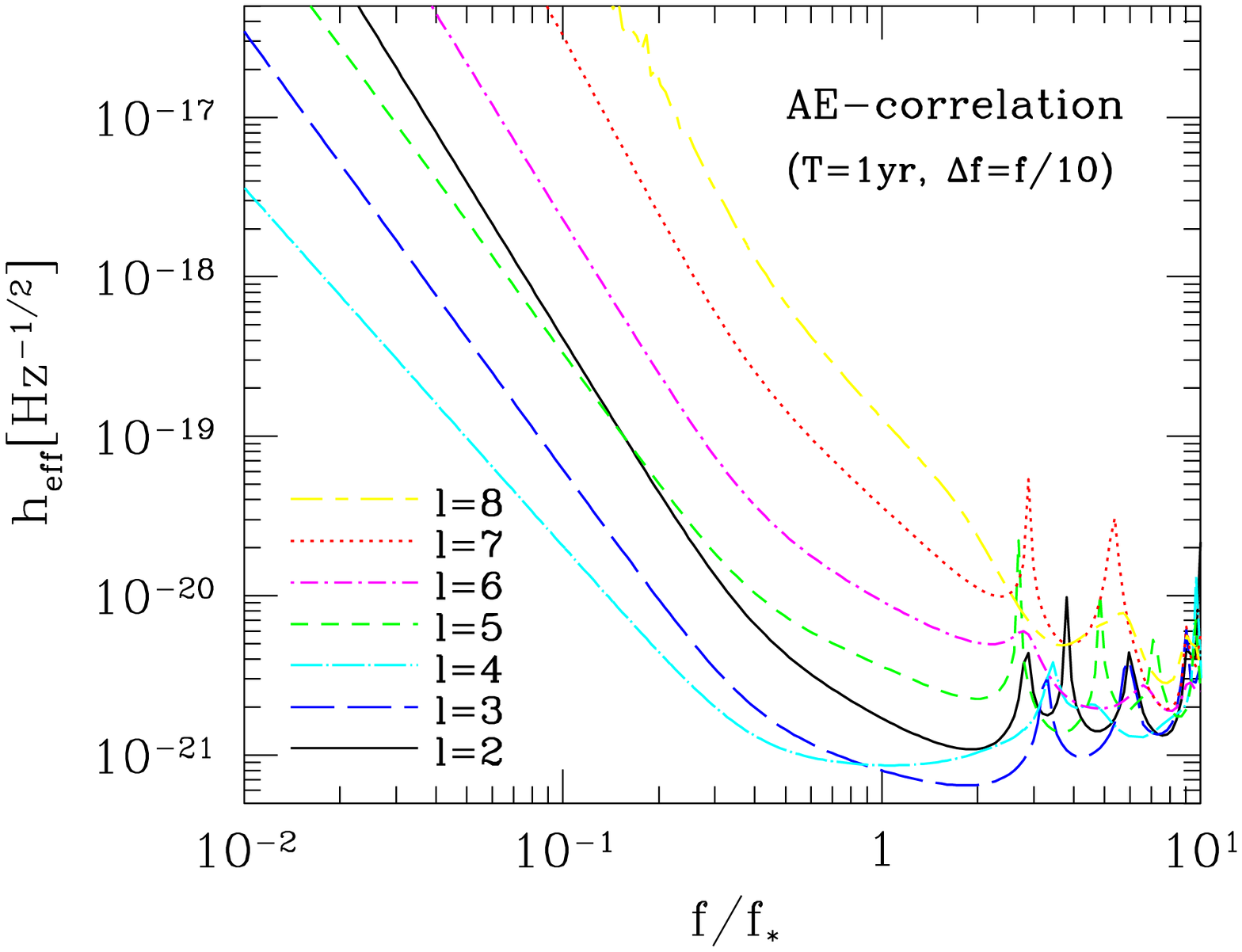}
\end{center}
\end{minipage}
&\hspace{0.5cm}
\begin{minipage}{70mm}
\vspace{5mm}
\begin{center}
\includegraphics[width=7cm,clip]{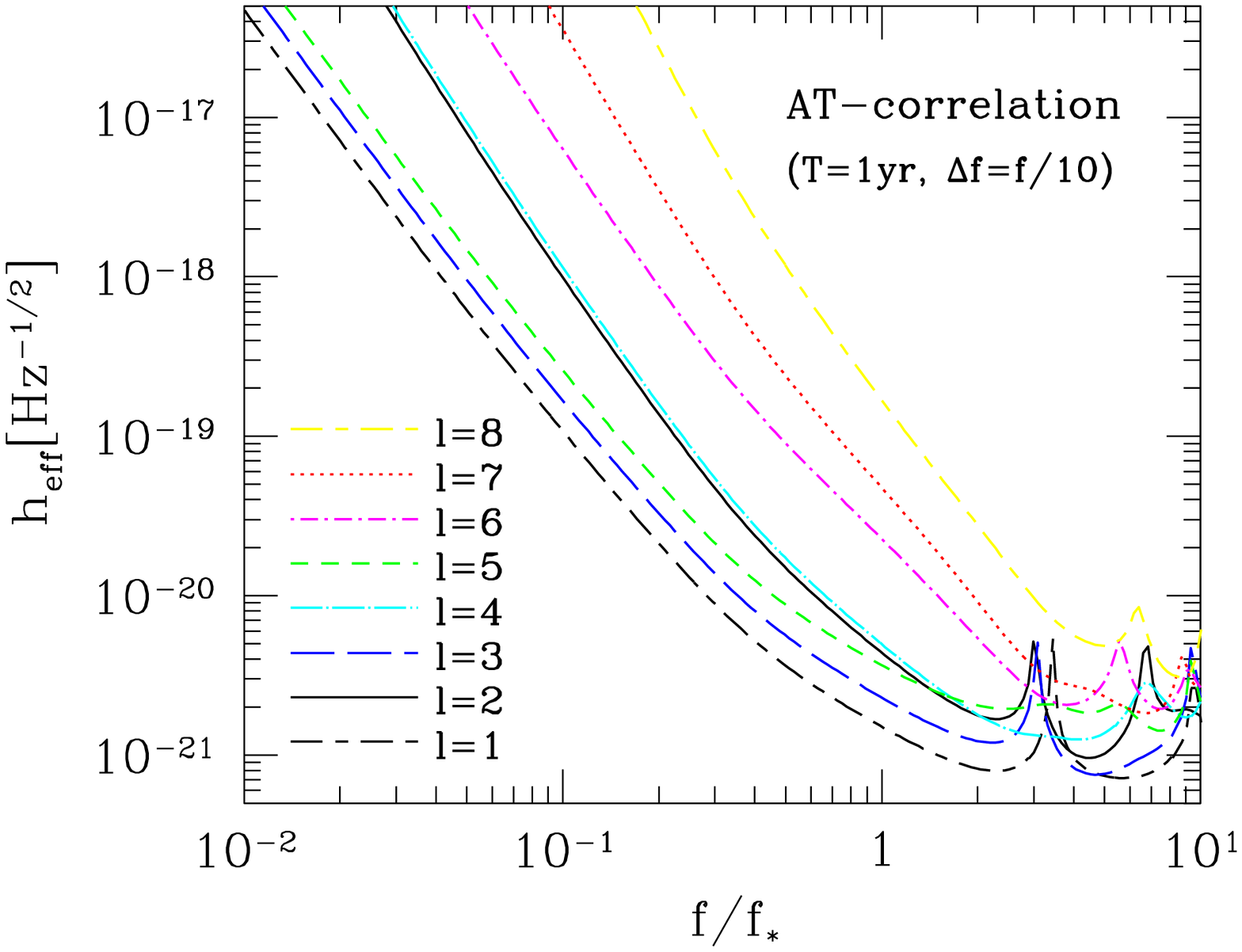}
\end{center}
\end{minipage}
\end{tabular}
\caption{
Effective sensitivity curves defined in 
(\ref{eq:h_eff_l_self}) and (\ref{eq:h_eff_l_cross}) for the 
self-correlated and cross-correlated optimal TDI variables: 
AA-correlation({\it top-left}); TT-correlation({\it top-right});  
AE-correlation({\it bottom-left}); AT-correlation({\it
 bottom-right}). In cases of the cross-correlation signals 
depicted in bottom panels, the observation time of $T=1$ year 
and the frequency resolution $\Delta f=f/10$ are assumed.  
The characteristic frequency is $f_* \simeq10$mHz. 
\label{fig:h_eff_optimal_TDIs}
}
\end{figure}
\end{center}

Figure \ref{fig:h_eff_optimal_TDIs} shows the effective sensitivity
curves for the self-correlated  and cross-correlated optimal TDIs as
functions of $\hat{f}=f/f_*$. 
In plotting these curves, we used the characteristic frequency 
$f_*\simeq10$mHz for the LISA detector. The different lines in each 
panel indicate the effective strain sensitivity for each multipole 
moment. Clearly, the directional sensitivity to a gravitational-wave 
background is not so good in the case of the self-correlated signals. 
As anticipated from Figure \ref{fig:sigma_optimal_TDIs} and the noise 
spectra (\ref{eq:noise_TT}), the effective sensitivity in the 
low-frequency limit scales as $h_{\rm eff}^{(\ell)}\propto f^{-2}$ for 
$\ell=0,\,2$ and $4$ of AA-correlation and  $h_{\rm eff}^{(\ell)}\propto
f^{-3}$ for $\ell=0,\,4$ and $6$ of TT-correlation. 
At the frequency around the characteristic frequency $f_*$, the 
directional sensitivities may reach at a maximal level and the 
higher multipole moment can be observed in both AA- and 
TT-correlations, however, the detectable multipole moments are 
still limited to $\ell=$even mode with 
$\ell\lesssim6$ for the sensitivity $h_{\rm eff}^{(\ell)}\sim 10^{-18}$ 
Hz$^{-1/2}$.

The situation might be improved if we consider the cross-correlation 
signals. In bottom panels of Figure \ref{fig:h_eff_optimal_TDIs},  
the observation time of $T=1$ year and the frequency resolution with 
interval $\Delta f=f/10$ are assumed. In this case, the sensitivity 
reaches $h_{\rm eff}^{(\ell)}\sim 5\times 10^{-21}$Hz$^{-1/2}$ in 
both AE- and AT-correlations (and also the ET-correlation), 
and the detectable multipole moments become, say, $\ell\lesssim10$ 
or even higher multipole moments in both $\ell=$odd and even modes. 
At the frequency $f/f_*\sim3$, the effective sensitivity for the 
higher multipole moments becomes comparable to that for the lower 
multipole and shows a complicated oscillatory behavior. 
Although the antenna pattern for cross-correlation signals is 
completely insensitive to the $\ell=0$ mode, improvement of the 
sensitivity is noticeable, which might be useful to distinguish 
between the gravitational-wave backgrounds from Galactic origin 
and those from extragalactic origin.

%
%
%

\section{Conclusion \& Discussions}
\label{sec:Conclusion}

In this paper, we discussed the directional sensitivity to
the anisotropy of gravitational-wave background observed
via space-based gravitational-wave detector. 
While the detection of anisotropic gravitational-wave background 
could be achieved utilizing the modulated signals of
cross-correlated data induced by the detector motion, the
directional sensitivity and the angular resolution crucially
depend on the antenna pattern function and/or the detector
response in detector's rest frame. In contrast to the groundbased
detector, the space interferometer with long baselines
gives a rather complicated response to the
gravitational-wave signals.

We have performed the spherical harmonic analysis of
antenna pattern function for space interferometer and
studied the general features of antenna pattern sensitivity
beyond the low-frequency approximation. We have shown
that the sensitivity to the multipole moments of an anisotropic
gravitational-wave background is generally restricted
by the geometry of the detector configuration and symmetries of the data combinations (see Table \ref{tab:summay_multipole}, 
Sec.~\ref{sec:Parity of detector response} and 
Sec.~\ref{subsec:geometric_relation}). 
The numerical analysis of the antenna pattern functions reveals that the angular power of the detector response increases with frequency and shows the complicated structures.  To characterize the directional sensitivity, we introduced the effective sensitivity 
$h_{\rm eff}^{(\ell)}(f)$ for each multipole moment and evaluated it in the case of the LISA detector specifically. 
Using the cross-correlated data of optimal TDIs, i.e., AE-, AT- and ET-correlations, we found that the detectable multipole moments with effective sensitivity $h_{\rm eff}^{(\ell)}\sim 10^{-20}$Hz$^{-1/2}$ may reach 
$\ell = 8 \sim 10$ or even higher multipoles at $f\sim f_*=10$ mHz, which would be useful to discriminate between the gravitational-wave backgrounds of Galactic origin and those of the extragalactic and/or the cosmological origins, recently discussed by several authors 
(e.g., \cite{Kosenko:1998,Schneider:2000sg,Farmer:2003pa}).

Although the improvement of the directional sensitivity beyond the low-frequency approximation is remarkable, the sensitivity of the space interferometer is still worse than the one achieved by the cosmic microwave background experiments, like the COBE (cosmic background explorer) and WMAP (Wilkinson microwave anisotropy probe). 
The one main reason is that the wavelength of the gravitational waves to which the space interferometer is sensitive is comparable to or longer than the arm length of the detector. 
Because of this, the response to the gravitational-wave background becomes simpler and most of the directional information is lost, as seen in Sec.\ref{sec:low_freq}. The directional sensitivity can be improved as increasing the frequency, however, the sensitivity beyond the characteristic frequency $f_*$ is limited by the instrumental noises. 
Another important aspect is that the phases of the gravitational-wave backgrounds are, in nature, random. Thus, the information of phase modulation induced by the detector motion cannot be used. 
This is marked contrast to the signals emitted from the point source, in which the angular resolution can reach at a level of a square degree or even better than that~
\cite{Cutler:1997ta,Moore:1999zw,Peterseim:1997ic,Takahashi:2003wm,Seto:2002uj}.

Further notice the important issues concerning the map-making capability of the gravitational-wave backgrounds. 
As discussed in Sec.\ref{sec:detection of anisotropy}, provided the time series data, the task is to solve the linear system (\ref{eq:deconvolution}) under a prior knowledge of the antenna pattern functions for the space interferometer. The crucial remark is that the antenna pattern functions for the cross-correlation signals taken from the optimal TDIs (A, E, T) are not independent 
(see Sec.\ref{subsec:geometric_relation}). 
This fact implies that Eq. (\ref{eq:deconvolution}) constructed from the three cross-correlation data (AE, AT, ET) is generally degenerate.
Thus, the deconvolution problem given in (\ref{eq:deconvolution}) would not be solved rigorously. 
Rather, one must seek a best-fit solution of $p_{\ell m}^E(f)$ from 
(\ref{eq:deconvolution}) under assuming a specific functional form of the luminosity distribution $p_{\ell m}^E(f)$.  
The analysis concerning this issue is now in progress and will be 
presented elsewhere. 
%
%
%
%
%
%
\begin{acknowledgments}
We would like to thank Y. Himemoto and T. Hiramatsu for valuable discussions and comments.
This work is supported by the 
Grant-in-Aid for Scientific Research of Japan Society for Promotion of Science (JSPS) (No.14740157).  
H.K. is supported by the JSPS. 
\end{acknowledgments} 
%
%
%
%
%
%
%
%
%
%
%
%
%
%
%
%
%
\appendix

\section{Spherical harmonic expansion in the low-frequency approximation}
\label{appendix:1}

In this appendix, employing the perturbative approach based on the 
low-frequency approximation $\hat{f}=f/f_*\ll1$, the spherical 
harmonic expansions for several antenna patterns are presented, 
which partially verify the properties summarized in Table 
\ref{tab:summay_multipole} and Eq. (\ref{eq:a_lm rule}). 
%
%
%
%
%
\subsection{Sagnac interferometers} 
%

The angular powers $\sigma_\ell$ of the self-correlated Sagnac 
signal $\mathcal{F}_{\scriptscriptstyle {\rm S}_i{\rm S}_i}$ and 
the cross-correlated 
Sagnac signal $\mathcal{F}_{\scriptscriptstyle {\rm S}_i{\rm S}_j}$ 
in the low-frequency approximation are summarized as follows: 
 \begin{eqnarray}
\mathcal{F}_{\scriptscriptstyle {\rm S}_i{\rm S}_i} :
& &
\sigma_{0} 
    =     \frac{4\sqrtpi\,{\hat{f}}^2  }{15}  
        - \frac{839\sqrtpi\, {\hat{f}}^4}{7560} , 
\quad
\sigma_{2}
=
    \frac{8 \sqrtpi\, {\hat{f}}^2  }{105}
    - \frac{32 \sqrtpi\, {\hat{f}}^4 }{945},
\quad
\sigma_{4}
=
    \frac{4 \sqrtpi\,{\hat{f}}^2  }{315} 
    - \frac{5233 \sqrtpi\,{\hat{f}}^4 }{997920} ,
\nonumber\\
&&
    \sigma_{6}
    =
    \frac{\sqrtpi\, }{432432}\sqrt{\frac{51641}{15}}  {\hat{f}}^4, 
\nonumber
\end{eqnarray}
\begin{eqnarray}
\mathcal{F}_{\scriptscriptstyle {\rm S}_i{\rm S}_j}:
&~&
    \sigma_{0}
    = \frac{2 \sqrtpi\, {\hat{f}}^2 }{15}
     - \frac{61 \sqrtpi\, {\hat{f}}^4 }{1080},
\quad
    \sigma_{1} = \frac{ \sqrtpi\, {\hat{f}}^3 }{ 126 },
\quad
    \sigma_{2}
     = \frac{4 \sqrtpi\,{\hat{f}}^2 }{105}
     - \frac{16 \sqrtpi\,{\hat{f}}^4}{945} ,     
\quad
    \sigma_{3} =
       \frac{\sqrtpi\, }{ 378} \sqrt{\frac{211}{10}} {\hat{f}}^3,
\cr
&~&
    \sigma_{4} 
    = \frac{\sqrtpi}{315} \sqrt{\frac{47}{3}} {\hat{f}}^2
    - \frac{12889 \sqrtpi}{166320 \sqrt{141} } {\hat{f}}^4,
\quad
   \sigma_{5} 
    = \frac{ \sqrtpi\,}{8316} \sqrt{\frac{689}{5}} {\hat{f}}^3,
\quad
   \sigma_{6} 
    = \frac{\sqrtpi\,}{4752} \sqrt{\frac{311}{1365}} {\hat{f}}^4.
\end{eqnarray}

%
%
%
%
\subsection{Optimal combinations of time-delay interferometry} 
%
From Eq. (\ref{eq:def AET mode}) the antenna pattern functions of the 
self-correlated optimal TDIs can be written down in terms of the 
antenna pattern for the Sagnac signals: 
\begin{eqnarray}
&& \mathcal{F}_{\scriptscriptstyle\rm AA}=
\frac{1}{2}\left\{
\mathcal{F}_{\scriptscriptstyle  3 3} 
+  \mathcal{F}_{\scriptscriptstyle 1 1} 
-  \mathcal{F}_{\scriptscriptstyle {(1,3)}} 
\right\},
\cr
&& \mathcal{F}_{\scriptscriptstyle\rm EE}=\frac{1}{{6}}\left\{
\mathcal{F}_{\scriptscriptstyle  1 1} 
+ 4\mathcal{F}_{\scriptscriptstyle 22}  
+ \mathcal{F}_{\scriptscriptstyle 33} 
+ \mathcal{F}_{\scriptscriptstyle {(1,3)}} 
-2 \mathcal{F}_{\scriptscriptstyle {(1,2)}} 
-2 \mathcal{F}_{\scriptscriptstyle {(2,3)}}  
  \right\},
\cr
&& \mathcal{F}_{\scriptscriptstyle\rm TT}=\frac{1}{ 3 }\left\{
\mathcal{F}_{\scriptscriptstyle  1 1} 
+ \mathcal{F}_{\scriptscriptstyle  2 2}  
+ \mathcal{F}_{\scriptscriptstyle 3 3} 
+ \mathcal{F}_{\scriptscriptstyle {(1,3)}} 
+ \mathcal{F}_{\scriptscriptstyle {(1,2)}} 
+ \mathcal{F}_{\scriptscriptstyle {(2,3)}} 
  \right\}.
\label{eq:Antenna, AA,EE,TT}
\end{eqnarray}
where the round brackets are an abbreviation for the symmetrization, 
$\mathcal{F}_{\scriptscriptstyle {(1,3)}} = 
\mathcal{F}_{\scriptscriptstyle {1}{3}} + 
\mathcal{F}_{\scriptscriptstyle {3}{1}} $ for instance, 
and 
$\mathcal{F}_{\scriptscriptstyle ij}$ stands for 
$\mathcal{F}_{\scriptscriptstyle {\rm S}_i {\rm S}_j}$. 
The antenna pattern functions of the cross-correlated signals, 
$\mathcal{F}_{\scriptscriptstyle\rm AE}$, 
$\mathcal{F}_{\scriptscriptstyle\rm AT}$ and 
$\mathcal{F}_{\scriptscriptstyle\rm ET}$, are 
\begin{eqnarray}
&& \mathcal{F}_{\scriptscriptstyle\rm AE}=\frac{1}{2\sqrt{3}}\left\{
\mathcal{F}_{\scriptscriptstyle 33} - 
\mathcal{F}_{\scriptscriptstyle 11} + 
\mathcal{F}_{\scriptscriptstyle 31} - 
\mathcal{F}_{\scriptscriptstyle 13} + 
2\mathcal{F}_{\scriptscriptstyle 12} -
2\mathcal{F}_{\scriptscriptstyle 32}   
  \right\},
\cr
&& \mathcal{F}_{\scriptscriptstyle\rm AT}=\frac{1}{\sqrt{6}}\left\{
\mathcal{F}_{\scriptscriptstyle 33} - 
\mathcal{F}_{\scriptscriptstyle 11} + 
\mathcal{F}_{\scriptscriptstyle 31} - 
\mathcal{F}_{\scriptscriptstyle 13} + 
\mathcal{F}_{\scriptscriptstyle 32} -
\mathcal{F}_{\scriptscriptstyle 12}   
  \right\},
\cr
&& \mathcal{F}_{\scriptscriptstyle\rm ET}=\frac{1}{3\sqrt{2}}\left\{
\mathcal{F}_{\scriptscriptstyle 11} - 
2\mathcal{F}_{\scriptscriptstyle 22} + 
\mathcal{F}_{\scriptscriptstyle 33} + 
\mathcal{F}_{\scriptscriptstyle 12}
- 2\mathcal{F}_{\scriptscriptstyle 21}
+ \mathcal{F}_{\scriptscriptstyle 32}
- 2\mathcal{F}_{\scriptscriptstyle 23}
+ \mathcal{F}_{\scriptscriptstyle {(1,3)}} 
  \right\}.
\label{eq:Antenna, AE,AT,ET}
\end{eqnarray}

Under the configuration in a specific coordinate 
(\ref{eq: a,b,c}), the low-frequency approximation of  
the multipole coefficients of the self-correlated optimal 
variables are given as follows: 
\begin{eqnarray}
\mathcal{F}_{\scriptscriptstyle\rm AA} : \quad &&
\cA_{{0 0}} =  \frac{2\sqrt{\pi}}{5} {\hat{f}}^2 ,
\quad  
\cA_{{2 0}} =  \frac{4}{7}\sqrt{\frac{\pi}{5}} {\hat{f}}^2 ,
\quad  
\cA_{{4 0}} =  \frac{ \sqrt{\pi}}{105} {\hat{f}}^2 ,
\quad  
\cA_{{4 4}} = - \left( \frac{1}{6}\sqrt{\frac{\pi}{70}} + 
\frac{i}{2}\sqrt{\frac{\pi}{210}} \right) 
{\hat{f}}^2,
\nonumber
\\
\mathcal{F}_{\scriptscriptstyle\rm EE} : \quad &&
\cA_{{0 0}} =  \frac{2\sqrt{\pi}}{5} {\hat{f}}^2 ,
\quad  
\cA_{{2 0}} =  \frac{4}{7}\sqrt{\frac{\pi}{5}} {\hat{f}}^2 ,
\quad  
\cA_{{4 0}} =  \frac{ \sqrt{\pi}}{ 105 }{\hat{f}}^2  ,
\quad   
\cA_{{4 4}} =  \left( \frac{1}{6}\sqrt{\frac{\pi}{70}} + 
\frac{i}{2}\sqrt{\frac{\pi}{210}} \right) 
{\hat{f}}^2,
\nonumber
\\
\mathcal{F}_{\scriptscriptstyle\rm TT} : \quad &&
\cA_{{0 0}} =  \frac{ \sqrt{\pi}}{ 504 }{\hat{f}}^4  ,
\quad   
\cA_{{4 0}} = - \frac{ \sqrt{\pi}}{1584}{\hat{f}}^4 ,
\quad   
\cA_{{6 0}} = - \frac{ \sqrt{\pi}}{11088\sqrt{13} }{\hat{f}}^4 ,
\quad   
\cA_{{6 6}} = - \frac{  \sqrt{\pi}}{ 48 \sqrt{ 3003} } {\hat{f}}^4 .
\end{eqnarray}
$m<0$ modes are given by the relation 
(\ref{eq:parity relation of a_l-m}). 
The rule (\ref{eq:a_lm rule}) strictly restricts the appearance of 
multipole moments, and of course the above multipole moments follow 
the rule (\ref{eq:a_lm rule}). 
For the cross-correlation of two data streams, one would expect that 
$\ell=$ odd modes appear even in the low-frequency limit. 
However it is not the case. A non-vanishing multipole moment is given 
to order $\mathcal{O} ({\hat{f}}^2 )$ by 
\begin{eqnarray}  
\mathcal{F}_{\scriptscriptstyle\rm AE} : \quad &&  
\quad
\cA_{{4 4}} =  \left( \frac{1}{2\sqrt{210}} -\frac{i}{6\sqrt{70}}  
\right)\sqrt{\pi} {\hat{f}}^2.
\end{eqnarray}
The $\ell=$ odd modes appear in the next order 
$ \mathcal{O} (\hat{f}^3)$ in some multipole moments that satisfy 
$\ell+m= \mathrm{even}$: 
\begin{eqnarray}
\mathcal{F}_{\scriptscriptstyle\rm AE} : \quad &&  
\cA_{{3 3}} =   \frac{1}{18} \sqrt{ \frac{7\pi}{15}}  {\hat{f}}^3, 
\quad 
\cA_{{5 3}} =     \frac{1}{18}\sqrt{\frac{\pi}{1155} } {\hat{f}}^3,
\end{eqnarray}
The angular powers up to $\mathcal{O}(\hat f^4)$ are summarized as  
\begin{alignat}{6}
\mathcal{F}_{\scriptscriptstyle\rm AA}, 
\mathcal{F}_{\scriptscriptstyle\rm EE} &:& 
\quad  
& \sigma_{0}  = \frac{2\sqrtpi}{5}{\hat{f}}^2 - 
        \frac{211\sqrtpi}{1260}{\hat{f}}^4, &
\quad  
& \sigma_{2} = \frac{4\sqrtpi}{35} {\hat{f}}^2
             - \frac{16\sqrtpi}{315}{\hat{f}}^4,&
\quad  
& \sigma_{4} = \frac{2\sqrtpi}{105} {\hat{f}}^2    
              - \frac{4511\sqrtpi}{498960}{\hat{f}}^4,& 
\quad  
& \sigma_{6} = \frac{\sqrtpi}{72072} \sqrt{\frac{1829}{15}} {\hat{f}}^4,& 
 \nonumber
\\
\mathcal{F}_{\scriptscriptstyle\rm TT} &:& \quad  
&\sigma_{0}  = \frac{ \sqrtpi }{504}  {\hat{f}}^4, &
\quad  
& \sigma_{4}  = \frac{\sqrtpi}{ 4752} {\hat{f}}^4 ,&
\quad  
 &\sigma_{6}  = \frac{\sqrtpi\,\sqrt{463} }{ 144144} {\hat{f}}^4&
\end{alignat}
and
\begin{eqnarray}
\mathcal{F}_{\scriptscriptstyle\rm AE} : \quad &&  
\sigma_{0} = 0, \quad
\sigma_{1} = 0, \quad
\sigma_{2} = \frac{{\sqrtpi\hat{f}^4}}{126\sqrt{3}}   ,\quad  
\sigma_{3} = \frac{{\sqrtpi\hat{f}^3}}{9\sqrt{30}}    ,
\nonumber\\
&&
\sigma_{4} =  \frac{\sqrtpi}{9\sqrt{ 35}}  {\hat{f}^2}
             -\frac{13\sqrtpi}{1188} \sqrt{\frac{5}{7}} {\hat{f}^4},\quad
\sigma_{5} = \frac{{\hat{f}}^3}{99\sqrt{210}}  ,\quad  
\sigma_{6} = \frac{{\sqrtpi\hat{f}^4} }{468\sqrt{210} } .   
\end{eqnarray}

%
%
%
%
\section{Parity transformation}   
\label{appendix:parity trans}         
%
%
%
%
%
%
%
Here, we summarize some formulae related to the parity transformation, 
which are used in Sec.\ref{sec:Parity of detector response}. 
In general, parity of the polarization tensor 
$\mathbf{e}^{+,\times}$ depends on the choice of the coordinate basis. 
Our choice of the basis vectors are those defined in 
(\ref{eq:polarization tensor}) and (\ref{eq: u,v,n}) just simply 
replacing the variables $\theta_E, \phi_E$ with 
$\theta, \phi$ in detector's rest frame. 
In the following, the vector ${\bf d}$ stands for the unit vectors 
${\bf a}, {\bf b}, {\bf c}$. We then obtain 
\begin{eqnarray}
{\mathbb{P}}\, ({\bf d} \cdot  {\bf \Omega}) &=& -1~({\bf d} 
\cdot{\bf \Omega}),  
\nonumber
\\
{\mathbb{P}}\, ({\bf d}\otimes{\bf d}):\mathbf{e}^A 
&=&  +1 ~ ({\bf d}\otimes{\bf d}):\mathbf{e}^A 
\quad (A= +,\times)
\end{eqnarray}
for the operator $\mathbb{P}$ and 
\begin{eqnarray}
{\mathbb{Q}}\, ({\bf d} \cdot  {\bf \Omega}) &=& + 1 ~({\bf d} 
\cdot{\bf\Omega}),  
\nonumber
\\
{\mathbb{Q}}\, ({\bf d}\otimes{\bf d}):\mathbf{e}^+ 
&=&  +1 ~ ({\bf d}\otimes{\bf d}):\mathbf{e}^+, 
\nonumber
\\
{\mathbb{Q}}\, ({\bf d}\otimes{\bf d}):\mathbf{e}^\times
 &=& - 1 ~ ({\bf d}\otimes{\bf d}): \mathbf{e}^\times
\end{eqnarray}
for the operator $\mathbb{Q}$. As for the composite operation 
$\mathbb{QP}$, which is identical to the parity transformation, one has 
\begin{eqnarray}
{\mathbb{Q\,P}}\, ({\bf d} \cdot  {\bf\Omega}) &=& 
-1~({\bf d} \cdot{\bf\Omega}), 
\nonumber
\\
{\mathbb{Q\,P}}\, ({\bf d}\otimes{\bf d}):\mathbf{e}^+ &=& 
 +1 ~ ({\bf d}\otimes{\bf d}):\mathbf{e}^+, 
\nonumber
\\
{\mathbb{Q\,P}}\, ({\bf d}\otimes{\bf d}):\mathbf{e}^\times
 &=& - 1 ~ ({\bf d}\otimes{\bf d}): \mathbf{e}^\times. 
\end{eqnarray}
%
%
%
%
%
%
%
%
%
\section{On cancellation of monopole and dipole moments in 
antenna pattern function for cross-correlated optimal TDIs}
\label{appendix:proof}
%
%
%
%
In this appendix, we will prove that the antenna pattern function for 
the cross-correlated optimal TDIs has the following symmetric properties: 
\begin{eqnarray}
&&   \sigma_0 (f) =0,\quad  \sigma_1 (f) =0 ~~~\mbox{for AE-correlation},
\cr
&&   \sigma_0 (f) =0 \quad \quad \quad \quad ~~~~~~~~
\mbox{for AT-,ET-correlation},   
\label{eq:sigma_AT_0}
\end{eqnarray}
which are intimately related to the geometric properties of both the 
detector configuration and the response function. 
As we have explained, the antenna pattern functions 
$\mathcal{F}_{\scriptscriptstyle\rm AE}$, 
$\mathcal{F}_{\scriptscriptstyle\rm AT}$ and 
$\mathcal{F}_{\scriptscriptstyle\rm ET}$ are 
written in terms of the antenna pattern functions for the Sagnac 
signals [see Eq.~(\ref{eq:Antenna, AE,AT,ET})].
The expressions readily imply that the multipole moments for the 
antenna pattern functions, 
$a_{\ell m}^{\scriptscriptstyle\rm AE}$,  
$a_{\ell m}^{\scriptscriptstyle\rm AT}$ and 
$a_{\ell m}^{\scriptscriptstyle\rm ET}$, are 
also obtained from the sum of the cross-correlated Sagnac signals, 
$a_{\ell m}^{\scriptscriptstyle {\rm S}_i{\rm S}_j}$ 
[see (\ref{eq:a_em ET, AT-Sij}), for example].

Let us first consider the monopole moment $a_{00}$. 
Since the monopole moment is obtained through the all-sky average of 
the antenna pattern function, it is, by construction, invariant under 
both the Euler rotation and the parity transformation of the coordinate 
system. 
This indicates that the monopole moments for various combinations of 
the Sagnac signals 
$a_{\ell m}^{\scriptscriptstyle {\rm S}_i{\rm S}_j}$ are degenerate 
and there are 
only two independent variables, that is, 
\begin{eqnarray}
a_{00}^{\scriptscriptstyle\rm S_1S_1}=
a_{00}^{\scriptscriptstyle\rm S_2S_2}=
a_{00}^{\scriptscriptstyle\rm S_3S_3}, 
\quad \quad
a_{00}^{\scriptscriptstyle{\rm S}_i{\rm S}_j}=
a_{00}^{\scriptscriptstyle{\rm S}_k{\rm S}_l}~~ (i\ne j,~k\ne l).  
\end{eqnarray}
Substituting this into the spherical harmonic expansion of the antenna 
pattern functions (\ref{eq:Antenna, AE,AT,ET}), we immediately see that 
the monopole component of cross-correlated optimal signals exactly 
vanishes, i.e., 
$a_{00}^{\scriptscriptstyle\rm AE}=
a_{00}^{\scriptscriptstyle\rm AT}=a_{00}^{\scriptscriptstyle\rm ET}=0$. 
Accordingly, the monopole moments of angular power, 
$\sigma_0^{\scriptscriptstyle\rm AE}$, 
$\sigma_0^{\scriptscriptstyle\rm AT}$ and 
$\sigma_0^{\scriptscriptstyle\rm ET}$, become vanishing.

Next focus on the dipole moment of AE correlation.  
In a specific choice of the coordinate system 
(\ref{eq: a,b,c}), all the components in the dipole moment vanish due 
to Eq. (\ref{eq:a_lm rule}) for the self-correlated Sagnac signals, 
i.e., $a_{1,m}^{\scriptscriptstyle {\rm S}_i{\rm S}_i}=0$.  
Also, the dipole moment with $m=0$ becomes zero for  
the cross-correlated signals, i.e., 
$a_{10}^{\scriptscriptstyle {\rm S}_i{\rm S}_j}=0$. 
Further, the relation (\ref{eq:parity relation of a_l-m}) implies 
$a_{1,-1}^{\scriptscriptstyle\rm AE}= 
[a_{11}^{\scriptscriptstyle\rm AE}]^*$. Collecting these facts, 
the dipole moment of angular power 
$\sigma_1^{\scriptscriptstyle\rm AE}$ can be written as 
\begin{equation}
\sigma^{\scriptscriptstyle\rm AE}_1
 = \sqrt{\frac{2}{3}}\,\,|a^{\scriptscriptstyle\rm AE}_{11}| = 
\frac{1}{3\sqrt{2}}\left|
a_{11}^{\scriptscriptstyle\rm  S_3S_1} -
a_{11}^{\scriptscriptstyle\rm  S_1S_3} +
2a_{11}^{\scriptscriptstyle\rm  S_1S_2} -
2a_{11}^{\scriptscriptstyle\rm  S_3S_2} \right|.
\label{eq:sigma_ell=1_AE}
\end{equation}
It is thus sufficient to consider the dipole moment with $m=+ 1$ for 
the cross-correlated Sagnac signals.

From Eq.~(\ref{eq:Rotation a_{em}-self-cross}), the angular power 
$\sigma_1^{\scriptscriptstyle\rm AE}$ can be solely determined by the quantity 
$a_{11}^{S_1S_2}$. 
If we write  $a_{11}^{\scriptscriptstyle\rm S_1S_2}
=r(\hat{f})\,e^{i\theta(\hat{f})}$, 
Eq. (\ref{eq:sigma_ell=1_AE}) becomes 
\begin{eqnarray}
 \sigma_1^{\scriptscriptstyle\rm AE}(\hat{f})
 =
 \frac{r }{3\sqrt{2}}\,\,
\left|2+e^{-i\,2\delta}+\left(1+2\,e^{-i\,2\delta}\right)\,
e^{-i\,2 \theta}\right|_{\delta=2\pi/3}.
\label{eq:reduced_sigma_1_AE}
\end{eqnarray}
To determine the phase factor $\theta(\hat{f})$ or amplitude 
$r(\hat{f})$, we recall the fact that the dipole moment of antenna 
pattern function $\mathcal{F}_{\scriptscriptstyle\rm AA}$ vanishes: 
\begin{eqnarray*}
a_{11}^{\scriptscriptstyle\rm AA}&=&\frac{1}{2}
\left(a_{11}^{\scriptscriptstyle\rm S_1S_1}
+a_{11}^{\scriptscriptstyle\rm S_3S_3}
-a_{11}^{\scriptscriptstyle\rm S_1S_3}
-a_{11}^{\scriptscriptstyle\rm S_3S_1}\right)
 = -\frac{1}{2} \left(a_{11}^{\scriptscriptstyle\rm S_1S_3}
+a_{11}^{\scriptscriptstyle\rm S_3S_1}\right)
\nonumber\\
 &=& 0.
\end{eqnarray*}
from (\ref{eq:def AET mode}).  Using the relations 
(\ref{eq:Rotation a_{em}-self-cross}), 
the above equation becomes
\begin{eqnarray*}
a_{11}^{\scriptscriptstyle\rm AA}=  \frac{r\,e^{i\,\theta}}{2} 
\left( e^{-i\, 2\theta}-e^{-i\, 2\delta}\right)_{\delta=2\pi/3}=0,   
\end{eqnarray*}
which finally yields $r(\hat{f})=0$ or $\theta(\hat{f})=\delta+n\pi$. 
Thus, substituting this value into the right-hand-side of equation 
(\ref{eq:reduced_sigma_1_AE}) immediately leads to the conclusion 
that the dipole moment of angular power 
$\sigma_1^{\scriptscriptstyle\rm AE}$ is exactly 
canceled. This means that all the dipole components 
for $a_{\ell m}^{\scriptscriptstyle\rm AE}$ become zero over the whole frequency range. 
%
%
%
%
\section{Coefficients in antenna pattern for toy model}
\label{appendix:toy model}
%
%
%
%
In this appendix, 
we summarize the coefficients $c^{(m,N)}$ and the functions 
$g_s^{(m,N)}(\hat{f})$ defined in (\ref{eq:func_gm}) and 
(\ref{eq:cal_I}).

For the coefficients $c^{(m,N)}$, the non-vanishing components are  
\begin{eqnarray}
&&c^{(0,0)}=-\frac{123\pi}{256},~~c^{(0,1)}=\frac{135\pi}{128},
~~c^{(0,2)}=-\frac{219\pi}{256},
\nonumber\\
&&c^{(2,0)}=-\frac{15\pi}{128},~~c^{(2,1)}=-\frac{39\pi}{128},
\nonumber\\
&&c^{(4,0)}=-\frac{9\pi}{512}.
\label{eq:c^{m,N}}
\end{eqnarray}
As for the functions $g_s^{(m,N)}(\hat{f})$, 
the non-vanishing components for $0\leq m\leq 4$ and $0\leq N\leq 2$ 
become 
\begin{eqnarray}
\begin{array}{llll}
g_0^{(0,0)}  =1,& 
g_0^{(2,0)}  =-4-\hat{f}^2,&
g_0^{(2,1)} = 2,& 
g_0^{(4,0)} = 144+48\hat{f}^2+\hat{f}^4,
\\
\\
g_1^{(2,0)} = 8i\,\hat{f},& 
g_1^{(2,1)} = -4i\,\hat{f},&
g_1^{(4,0)} = -576i\,\hat{f}-32i\,\hat{f}^3,&
\\
\\
g_2^{(0,1)}  =1,& 
g_2^{(2,0)}  =12+2\hat{f}^2,&
g_2^{(2,1)}  =-24-\hat{f}^2,&
g_2^{(4,0)}  =-1440-432\hat{f}^2 -4\hat{f}^4,
\\
\\
g_3^{(2,0)}  =-8i\,\hat{f},& 
g_3^{(2,1)}  = 16i\,\hat{f},&
g_3^{(4,0)}  = 1920i\,\hat{f} +96i\,\hat{f}^3,&
\\
\\
g_4^{(0,2)}  =1,& 
g_4^{(2,0)}  =-\hat{f}^2,&
g_4^{(2,1)}  =30+2\hat{f}^2,&
g_4^{(4,0)} =1680+720\hat{f}^2 +6\hat{f}^4,
\\
\\
g_5^{(2,1)} = -12i\,\hat{f},& 
g_5^{(4,0)} = -1344i\,\hat{f} - 96i \hat{f}^3,&&
\\
\\
g_6^{(2,1)} =-\hat{f}^2,& 
g_6^{(4,0)} =-336\hat{f}^2 - 4\hat{f}^4,&&
\\
\\
g_7^{(4,0)} = 32i\,\hat{f}^3,&&& 
\\
\\
g_8^{(4,0)} =\hat{f}^4.&&& 
\end{array}
\nonumber
\end{eqnarray}
Note that the other components with $m\geq0$ do not contribute 
to the calculation of the multipole coefficient $a_{\ell m}$ 
due to the coefficients $c^{(m,N)}$ (Eq. (\ref{eq:c^{m,N}})).
%
%
%
%
%
%
%
%
%
%

\bibliographystyle{apsrev} 


\end{document}